\documentclass[]{aastex}

\usepackage{emulateapj5}
\usepackage{apjfonts}

\slugcomment{To appear in PASP}

\def\dfrac#1#2{{\displaystyle\frac{#1}{#2}}}
\def\ex#1{{\sf E}[#1]}
\def\var#1{{\sf V}[#1]}
\makeatletter
\def\lnyoro{\mathrel{\mathpalette\gl@align<}}
\def\gnyoro{\mathrel{\mathpalette\gl@align>}}
\def\gl@align#1#2{\lower.6ex\vbox{\baselineskip\z@skip\lineskip\z@\ialign{$\m@th#1\hfil##\hfil$\crcr#2\crcr\sim\crcr}}}
\makeatother
\def\iso{{\sl ISO}}

\def\iras{{\sl IRAS}}
\def\dl{d_{\rm L}}
 
\begin{document}

\title{Impact of Future Submillimeter and Millimeter Large Facilities on 
the Studies of Galaxy Formation and Evolution}

\author{Tsutomu T. Takeuchi\altaffilmark{1},
  Ryohei Kawabe\altaffilmark{2}, 
  Kotaro Kohno\altaffilmark{2},
  Koichiro Nakanishi\altaffilmark{2,5}\\
  Takako T. Ishii\altaffilmark{3}
  Hiroyuki Hirashita\altaffilmark{4,5}, 
  and
  Kohji Yoshikawa\altaffilmark{4,5}
}

\bigskip

\altaffiltext{1}
{Division of Particle and Astrophysical Sciences, Nagoya University,
  Chikusa-ku, Nagoya, 464--8602, JAPAN.}
\altaffiltext{2}
{Nobeyama Radio Observatory, Minamimaki, Minamisaku, Nagano, 
  384--1305, JAPAN.}
\altaffiltext{3}
{Kwasan and Hida Observatories, Yamashina-ku, Kyoto University, 
  Kyoto, 607--8471, JAPAN.}
\altaffiltext{4}
{Department of Astronomy, Faculty of Science, Kyoto University,
  Sakyo-ku, Kyoto 606--8502, JAPAN.}
\altaffiltext{5}
{Research Fellows of the Japan Society for the Promotion of
  Science.}
\affil
{E-mail: takeuchi@u.phys.nagoya-u.ac.jp; 
  kawabe, kotaro, nakanisi@nro.nao.ac.jp;
  ishii@kwasan.kyoto-u.ac.jp;
  hirasita, kohji@kusastro.kyoto-u.ac.jp; 
}

\begin{abstract}

We investigate what we can learn about galaxy formation and 
evolution from the data which will be obtained by the forthcoming 
large submillimeter/millimeter facilities, mainly by the Atacama Submillimeter
Telescope Experiment (ASTE) and the Atacama Large Millimeter Array/Large 
Millimeter and Submillimeter Array (ALMA/LMSA).
We first calculate the source counts from $350\;\mu$m to 3~mm using
the empirical infrared galaxy number count model of Takeuchi et al.\ (2001).
Based on the number counts, we evaluate the source confusion 
and determine the confusion limit at various wavebands 
as a function of the characteristic beam size.
At submillimeter wavelengths, source confusion with the 10 -- 15-m class 
facilities becomes severe at 0.1 to 1~mJy level, 
and astrometry and flux measurements will be difficult.
However, we show that very a large-area survey of submillimeter sources 
brighter than 10 -- 50~mJy can provide a unique constraint on infrared 
galaxy evolution at $z = 1\mbox{--}2$, and such a survey is suitable for 
the ASTE.
In addition, such a survey enables us to study the clustering properties of 
submillimeter sources, which is still highly unknown.
We also find that the $5\sigma$-confusion limit of the LMSA is 
fainter than $1 \; \mu$Jy, which enables us to study the contribution of 
sources at extremely large redshift.
When we discuss such a deep flux limit, the dynamic range of a detector 
should be taken into account, since extremely bright sources make it 
impossible to detect the faintest sources near the detection limit.
We evaluate the probability that sources that are $10^3$ times brighter 
than the $5\sigma$-detection limit of the LMSA and the ALMA exist in the 
field of view.
We find that the probability is $\lnyoro 3 \times 10^{-4}$, and therefore 
we do not have to worry about the dynamic range.
The source counts at such faint flux levels give important information of 
the epoch of galaxy formation.
We then show that multiband photometry from the infrared (by ASTRO-F) to 
the millimeter can be utilized as a redshift estimator.
We examined the performance of this method by Monte Carlo simulations and 
found that it successfully works if we have reasonable measurement accuracy.
In addition, we compare the observed 1.4, 5, and 8-GHz source counts with
our model counts to examine the contribution of star forming galaxies to 
the faint radio galaxies.
We find that the faintest radio number counts ($\sim 1\; \mu {\rm Jy}$) 
are dominated by actively star-forming galaxies which lie at intermediate
redshift $z \sim 1 \mbox{--}2$.
\end{abstract}


\section{INTRODUCTION}\label{sec:introduction}

A significant fraction of galaxies in their youth are expected to be dusty 
starburst since the first epoch of star formation produces a lot of dust.
The radiation of underlying stars is absorbed by dust that re-radiates in 
the far infrared (IR) (for a review, see e.g., Sanders \& Mirabel 1996).
For distant galaxies, this emission is redshifted to the the submillimeter 
and millimieter wavelengths.
Hence, observations at these wavelengths are regarded as key for the studies 
of the formation and early evolution of galaxies.

The first product of any galaxy survey is the number counts, i.e.\ the number 
of galaxies as a function of limiting flux in a certain sky area.
The evolution of galaxies is often investigated through number counts.
Number counts alone do not provide non-degenerate information on the 
redshifts of detected sources; therefore one constructs a suitable model and 
extracts the history of galaxy evolution through the model.
Various models of IR galaxies have been proposed
(e.g., Beichman \& Helou 1991; Franceschini et al. 1994; 
Pearson \& Rowan-Robinson 1996; Guiderdoni et al. 1998; Malkan \& Stecker
1998; Tan, Silk, \& Balland 1999; Takeuchi et al. 1999, 2001; and 
Pearson 2000) to understand and interpret the observational results.
Recent studies using the Infrared Space Observatory (\iso) 
(e.g., Kawara et al.\ 1998; Puget et al.\ 1999; Dole et al.\ 2000; 
Serjeant et al.\ 2000; Efstathiou et al.\ 2000; Kawara et al.\ 2000; 
and Okuda 2000)
and the Submillimeter Common-User Bolometer Array (SCUBA) 
(e.g., Smail et al.\ 1997; Barger et al.\ 1998; Hughes et al.\ 1998; 
Barger, Cowie, \& Sanders 1999; Eales et al.\ 1999, 2000; 
Holland et al.\ 1999; and Blain et al.\ 1999, 2000) have stimulated this field.
Unfortunately, observations in the submillimeter wavebands are difficult, 
mainly because of the following reasons.
First, ground-based observations in these wavelengths are hard because 
of the atmospheric absorption.
Second, state-of-the-art instruments are required both for the 
antennas and for the detectors.
New breakthroughs in the submillimeter field are just around 
the corner with the Large Millimeter and Submillimeter Array (LMSA) and 
the Atacama Large Millimeter Array (ALMA).

The LMSA\footnote{URL: {\sf http://www.nro.nao.ac.jp/\~{}
lmsa/index.html.}} 
is the ground-based radio interferometric facility proposed by Japan.
In the current design concept the array will consist of 32 12-m antennas, 
whose total collecting area is $3619\;{\rm m}^2$.
The baselines can be $20\;{\rm m}$ to $10000\;{\rm m}$, which realizes
the maximum angular resolution of $0.01\; {\rm arcsec}$ at 345 GHz.
The receivers will cover observing frequencies from 80 to 900 GHz. 
The array will be located at Pampa la Bola/Llano de Chajnantor, a very high 
site in Chile, to realize sub-arcsec resolution imaging at the very high 
frequencies. 
The Atacama Large Millimeter Array (ALMA)\footnote{URL: 
{\sf http://www.mma.nrao.edu/} (U.S.\ side) and 
{\sf http://www.eso.org/projects/alma/} (European side).} 
is the new name for the merger of the major millimeter array projects 
into one global project: 
the European Large Southern Array (LSA), the U.S.\ Millimeter Array (MMA), 
and the LMSA. 
Hereafter we use the ALMA to describe the whole.
The $5\sigma$ sensitivities of the LMSA at wavelengths 350 $\mu$m, 450 $\mu$m, 
650 $\mu$m, 850 $\mu$m, 1.3 mm, and 3.0 mm (mean values in winter season)
are 1200, 660, 370, 49, 23, and 14 $\mu$Jy beam$^{-1}$, respectively
(8-hour integration and 8-GHz bandwidth) in dual polarization and DSB mode.
For the reference of the atmosphere condition and system temperature 
$T_{\rm sys}$, see Matsushita et al.\ (1999).
The ALMA will consist of 96 12-m antennas, and in this case
the $5\sigma$ sensitivities at the above wavelengths 
are 390, 220, 120, 16, 7.5, and 4.6 $\mu$Jy beam$^{-1}$, respectively, 
under the same assumptions.

The Atacama Submillimeter Telescope Experiment (ASTE) is a Japanese project 
to install and operate a 10-m submillimeter antenna at Pampa la Bola in the 
Atacama desert in northern Chile. 
A new 10-m antenna has been constructed in Nobeyama Radio Observatory in
February 2000, and is scheduled to be installed in Chile in 2001.
The telescope will be equipped with submillimeter-wave SIS mixer receivers
and a submillimeter-wave direct detector camera.
This project has two goals:
One is to examine the 10-m submillimeter antenna as the technical prototype of 
the LMSA at the site of northern Chile, and the other is to explore the 
possibility of the submillimeter observations in the southern hemisphere and 
to obtain astrophysical results.
Details of the present development of the ASTE is given in e.g., Matsuo et 
al.\ (2000) and Ukita et al.\ (2000)\footnote{We should also mention that 
the SMA (Sub-Millimeter-wave Array) will be operational with all 8 telescopes 
on Mauna Kea, Hawaii, by the end of 2001.}.

Before the ALMA/LMSA begins operation, a Japanese infrared satellite 
ASTRO-F will be launched\footnote{
URL: {\sf http://www.ir.isas.ac.jp/ASTRO-F/index-e.html}.}.
The ASTRO-F will carry out a FIR all-sky survey at two narrow bands, N60 
($50 - 70\;\mu$m) and N170 ($150 - 200\;\mu$m) and two wide bands, WIDE-S 
($50 - 110\;\mu$m) and WIDE-L ($110 - 200\;\mu$m).
The detection limits are estimated as 39~mJy and 110~mJy for N60 and N170, 
and 16~mJy and 90~mJy for WIDE-S and WIDE-L, respectively 
(Takahashi et al.\ 2000). 
A huge database of IR galaxies can be expected from the survey.
Some other stratospheric and space facilities such as SOFIA 
(Becklin 1998), {\sl SIRTF} (Fanson et al. 1998), and {\sl FIRST} 
(Pilbratt 1998) are also in progress, but we stress that the ASTRO-F is the 
one and only all-sky surveyor among these facilities\footnote{Since the 
following discussion of this paper is based on large survey-type datasets, 
we hereafter concentrate on the estimation based on the ASTRO-F, ASTE, and 
ALMA/LMSA.
Similar discussion for other facilities can be done straightforwardly.}.
The combination of the FIR data from the ASTRO-F and the submillimeter and 
millimeter data from the ASTE and the ALMA/LMSA promises to a new era
of extragalactic astrophysics and observational cosmology.

In this paper, we study and examine what we can learn about galaxy formation 
and evolution from the unprecedentedly high-quality data which will be 
obtained by the ASTE and the ALMA/LMSA through a simple empirical 
galaxy number count model proposed by Takeuchi et al.\ (2001).
The rest of the paper is organized as follows. 
In Section~\ref{sec:model} we present the IR galaxy number count model
on which our subsequent discussions are based.
We also formulate the source confusion limit and optimal survey area
to suppress both the fluctuation caused by the source clustering, and
the Poisson shot noise by the sparseness of the sources.
In Section~\ref{sec:result} we show the galaxy number counts at submillimeter
and millimeter wavelengths and extensively examine how galaxy evolution 
and formation epoch affect the number counts.
We also discuss the optimal strategy of a survey at these wavelength
to obtain the maximal information on galaxy evolution and formation.
Section~\ref{sec:conclusion} is devoted to our summary and conclusions.

\section{MODEL DESCRIPTION}\label{sec:model}

\subsection{Number Count Model}\label{subsec:NC}

The galaxy number count model is represented by galaxy spectral energy 
distributions (SEDs), luminosity function (LF), cosmology, and 
galaxy evolution.
We briefly review the number count model by Takeuchi et al.\ (2001; hereafter
T01).
Based on this model, we calculate galaxy number counts with various galaxy 
evolution histories in the subsequent part of this paper.

\begin{figure*}[t]
\centering\includegraphics[width=9cm]{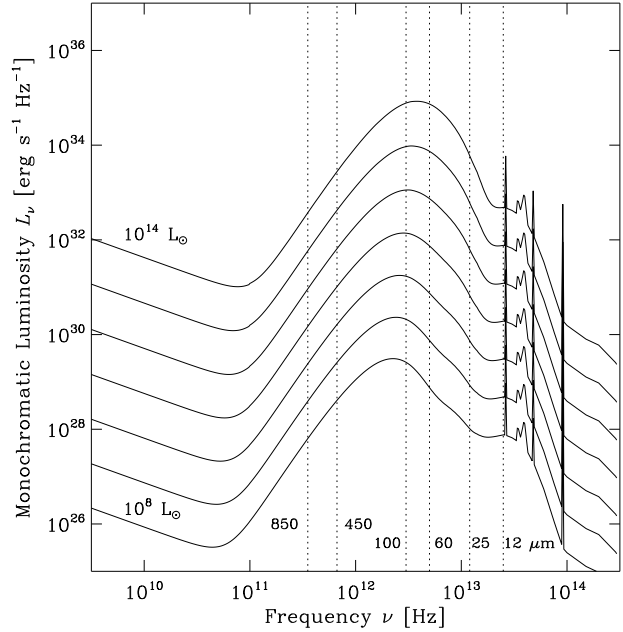}
\figurenum{1}
\figcaption[fig1.ps]{
  The assumed galaxy spectral energy distribution (SED) in the near infrared 
  to the radio wavelengths.
  The prominent emission bands are PAH features.
  The SEDs with the FIR luminosity of $10^8$, $10^9$, $10^{10}$, $10^{11}$, 
  $10^{12}$, $10^{13}$, and $10^{14}\;L_\odot$ are shown from the 
  bottom in this order.
  These construction of these SEDs are described by Takeuchi et al. (2001).
  The vertical dotted lines depict the wavebands of {\sl IRAS} and the
  two representative wavelengths of the atmosperic window in the 
  submillimeter.
}\label{fig:sed}
\end{figure*}

The FIR SED is constructed based on the {\sl IRAS}~color--luminosity relation 
(Smith et al. 1987; Soifer \& Neugebauer 1991),
\begin{eqnarray}\label{eq:color_flux}
  \log {\displaystyle\frac{S_{60}}{S_{100}}} = (0.10 \pm 0.02) \log L_{60} 
  - (1.3 \pm 0.2)\;,
\end{eqnarray}
where $S_\lambda$ is the detected flux density at wavelength 
$\lambda$ [$\mu$m], and $L_{60}$ [$L_\odot$] is the intrinsic luminosity
evaluated at $60$-$\mu$m, $L_{60} \equiv \nu L_\nu$.
This relation is converted to the dust temperature $T_{\rm dust}$--$L_{60}$ 
relation, then the modified blackbody continuum with the corresponding 
$T_{\rm dust}$ are calculated.
We then added the mid-IR and radio continuum spectrum based on the empirical
correlations between fluxes in these wavelengths and FIR continuum.
The final SEDs that we use in our number count and CIRB models are presented
in Figure~\ref{fig:sed}.
For the full description of the SED construction, see T01.

The $60$-$\mu$m LF based on {\sl IRAS}~data by Soifer et al.~(1987)
is adopted as the local IR LF of galaxies and pure luminosity evolution 
is assumed in this study.
Thus the 60-$\mu$m luminosity of a galaxy at redshift $z$
is described by
\begin{eqnarray}
  L_{60} (z) = L_{60} (0) f(z)\;. \label{deflev}
\end{eqnarray}
We also assumed that the luminosity evolution is `universal', i.e.\ 
its functional form is fixed. 
In this paper we call $f(z)$ `the evolutionary factor' as in T01.

Using the above formulae, we calculate the galaxy number counts.
As we see later, we set the maximum redshift we consider the counts of 
galaxies into the calculation, $z_{\rm max} = 5$ and we vary it 
from $z_{\rm max} = 2$ to $z_{\rm max} = 7$ to investigate
the effect of $z_{\rm max}$ on the number counts.
T01 statistically estimated the permitted range of the
evolutionary factor from the observed IR number counts and the CIRB spectrum.
The evolutionary factor is directly connected to the evolution of the
cosmic IR luminosity density, which is closely related to the cosmic star
formation history (e.g., Kennicutt 1998).
The conversion from the IR luminosity to the star formation rate and the 
interpretation of the evolutionary factor to the cosmic star formation history
are discussed in T01.
In this paper we examine how the number counts at submillimeter and millimeter
wavelengths vary with different evolutions, based on the two 
representative evolutions from T01 (we call them Evolution 1, and 2 in 
this paper).

\begin{figure*}[t]
\centering\includegraphics[width=8cm]{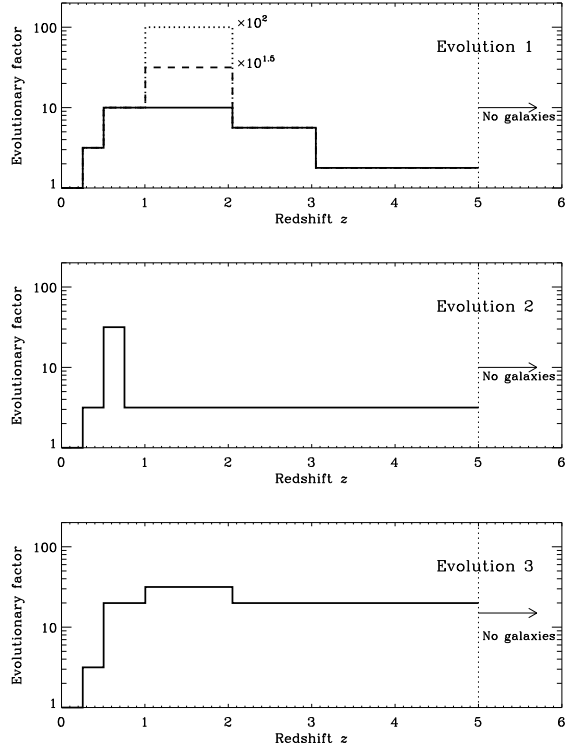}
\figurenum{2}
\figcaption[fig2.ps]{
  The three representative evolutionary history of galaxies used in this study.
  The upper and middle models are estimated from the infrared galaxy 
  number counts (Takeuchi et al. 2001).
  The bottom model is the new additional one constructed to reproduce
  the observed submillimeter galaxy number counts.
  In the top panel, dashed and dotted lines represent the modified 
  evolutionary factors to examine the effect on the evolution at 
  $z = 1 \sim 2$ to the number count at submillimeter.
}\label{fig:evlp}
\end{figure*}

The model of T01 has been constructed based on the properties of IR galaxies, 
and it slightly underestimates the submillimeter source counts.
Therefore in this study, we use an additional model based on the SCUBA source 
counts.
The additional evolution model overestimates the CIRB spectrum, but 
the possibility of reconciling the CIRB and submillimeter number counts
is discussed in Takeuchi et al.\ (2000); therefore, we do not go into the 
details of this problem further.
We call this model `Evolution 3'.
These three evolutions are summarized in Figure~\ref{fig:evlp}.

\subsection{Formulation of Source Confusion}

We summarize the framework of the source confusion statistics and 
derive the source confusion limit.
The basic formulation of the source confusion was first discussed by
Scheuer (1957).
The formula for the confusion limit for power-law number counts was
formulated in the classical work of Condon (1974), and extended for 
the case of general number counts by Franceschini et al.\ (1989).

In this subsection, we briefly re-formulate these issues
and examine the formulae carefully for precise numerical calculations.
The original formulation of Scheuer (1957) was rather complex, so
we use a simper but modern theory of random summation
(e.g., Gnedenko \& Korolev 1996) to clarify the mathematical structure 
of the problem.

\subsubsection{Statistics of Signal from Faint Sources in the Beam}

First we define a differential number count per $\mbox{sr}$ with
flux $S \in [S, S + d S]$, $n(S)\,d S$, 
\begin{eqnarray}
  n (S) \, d S = \left. \dfrac{d N}{d S}\right|_S d S \;,
\end{eqnarray}
where $N = N (>S)$ is the cumulative number count.
As a radio telescope scan across the sky, the antenna temperature 
fluctuates due to many confusing sources passing through the beam.
Let $h(\theta, \phi)$ be the beam pattern (normalized to unity
at the beam center), and $x = Sh(\theta, \phi)$, the response of the 
telescope to a source of flux density $S$ located at an angular position
$(\theta, \phi)$ from the beam axis.
The mean number of source responses of intensity $x$ in a beam is 
\begin{eqnarray}\label{eq:diff_nc_beam}
  R(x) d x = \int_{\Omega_{\rm beam}} 
  n \left( \dfrac{x}{h(\theta, \phi)} \right) 
  \dfrac{d x}{h(\theta, \phi)} \, d \Omega \; .
\end{eqnarray}
The amplitude of the total signal at any point, $D$, is a 
randomly-stopped sum of the responses $x$ due to all the sources in the 
response pattern:
\begin{eqnarray*}
  D = x_1 + x_2 + \cdots + x_K\; ,
\end{eqnarray*}
where $K$ is the total number of contributing sources.
We should note that $K$ randomly varies and takes an integer value.
The mean number of $x$'s, $\kappa$, is 
\begin{eqnarray}
  \kappa = \int^{\infty}_{-\infty} R(x) d x  = 
  \int^{\infty}_{0} R(x) d x \; .\nonumber
\end{eqnarray}
The probability that the total number of $x$'s has a value $k$ is 
given by a Poisson distribution
\begin{eqnarray}\label{eq:poisson}
  p_k = \dfrac{\kappa^k}{k!} e^{-\kappa}\; .
\end{eqnarray}
We have the (cumulative) distribution function (d.f.) of $x$, $F(x)$, 
\begin{eqnarray}\label{eq:dfx}
  F(x) = {\sf P}(x' < x) 
  = \int^{x}_{-\infty} \dfrac{R(x')}{\kappa} d x' \; .
\end{eqnarray}
Then the d.f.\ of a signal $D$ with $k$ summands is 
\begin{eqnarray}
  F^{*(k+1)}(D) &=& \int^{D}_{0} F^{*k}(D-x')\, d F(x') \; ,
  \label{eq:dfx_conv} \\
  F^{*1}(D) &=& F(D) \; .\label{eq:dfx_init}
\end{eqnarray}
Combining eqs.~(\ref{eq:poisson}), (\ref{eq:dfx}), (\ref{eq:dfx_conv}), and 
(\ref{eq:dfx_init}), we obtain the d.f.\ of $D$, $G(D)$:
\begin{eqnarray}
  G(D) = \sum_{k=1}^{\infty} p_k F^{*k}(D)\;. \nonumber
\end{eqnarray}
The characteristic function (c.f.) of $G(D)$, $\Phi_G(t)$, is expressed
by the c.f.\ of $F(D)$, $\Phi_F(t)$, as 
\begin{eqnarray}
  \Phi_G(t) = \int_{-\infty}^{\infty} e^{itD} dG(D)
  = \sum_{k=1}^{\infty} p_k \Phi_F(t)^k  \;.
\end{eqnarray}
Recall that $p_k = (\kappa^k e^{-\kappa})/k!$, then we have
\begin{eqnarray}
  \Phi_G(t) &=&
  \sum_{k=1}^{\infty} \dfrac{\kappa^k e^{-\kappa}}{k!} \Phi_F(t)^k  \nonumber\\
  &=& 
  e^{-\kappa} \sum_{k=1}^{\infty} \dfrac{\kappa^k \Phi_F(t)^k}{k!} \nonumber\\
  &=& 
  e^{-\kappa} e^{\kappa\Phi_F(t)} \nonumber = 
  e^{\kappa (\Phi_F(t) -1)}\; .
\end{eqnarray}
The expectation value and variance of a randomly-stopped sum $D$ are 
known to be 
\begin{eqnarray}
  \ex{D} &=& \ex{\,k\,}\,\ex{\,x_1} \; , \nonumber \\
  \var{D} &=& \ex{\,k\,}\,\var{\,x_1} + \var{\,k\,}(\ex{\,x_1})^2 \nonumber \\
\end{eqnarray}
(e.g., Stuart \& Ord 1994) where $\ex{\;\cdot\;}$ and $\var{\;\cdot\;}$ 
represent the expectation and variance, respectively.
Thus we have 
\begin{eqnarray}
  \ex{D} &=& \ex{\,k\,}\,\ex{\,x\,} = \kappa \,\ex{\,x\,}  \nonumber \\
  &=& \kappa \int_{-\infty}^{\infty} x \frac{R(x)}{\kappa} d x \nonumber\\
  &=& \int_{-\infty}^{\infty} x R(x) d x  \; , \\
  \var{D} &=& \ex{\,k\,}\,\var{\,x\,} + \var{\,k\,}(\ex{\,x\,})^2 \nonumber \\
  &=& \kappa (\ex{(x - \ex{\,x\,})^2} + (\ex{\,x\,})^2) \nonumber \\
  &=& \kappa \ex{\,x^2} = \int_{-\infty}^{\infty} x^2 R(x) d x  \;.
\end{eqnarray}

\subsubsection{Confusion limit}

We next formulate the relation between the beam size $\theta_0$ and the
source confusion limit.
Hereafter we assume that $h(\theta, \phi) = h(\theta)$.
A general case without this assumption is considered by Condon (1974),
but the result is not substantially affected by the above simplification.
First we discuss the case that the number count is described by a power-law:
\begin{eqnarray}
  n(S) = \alpha S^{-\gamma}\; . 
\end{eqnarray}
Then we have
\begin{eqnarray}
  n \left( \dfrac{x}{h(\theta)} \right) 
  = \alpha \left( \dfrac{x}{h(\theta)} \right)^{-\gamma} 
  = \alpha h(\theta)^{\gamma} x^{-\gamma}\; .
\end{eqnarray}
The mean number of $x$, $R(x)$ is
\begin{eqnarray}
  R(x) = \int_{\Omega_{\rm beam}} \alpha h(\theta)^{\gamma} x^{-\gamma}
  \dfrac{d \Omega}{h(\theta)}
  = \alpha\Omega_{\rm eff} x^{-\gamma}\; ,
\end{eqnarray}
where $\Omega_{\rm eff}$ is
\begin{eqnarray}
  \Omega_{\rm eff} \equiv
  \int_{\Omega_{\rm beam}} h(\theta)^{\gamma - 1} d \Omega \; .
\end{eqnarray}
Here we consider the Gaussian beam pattern: 
\begin{eqnarray}\label{eq:gaussian_beam}
  h(\theta) = e^{-(4\ln 2)\left(\frac{\theta}{\theta_0}\right)^2}\;.
\end{eqnarray}
We note that more complex beam patterns can be taken into account.
Other beam patterns have been commented upon by e.g., Scheuer (1957) and 
Franceschini et al.~(1989).
But in a real situation, the result does not depend on the details of the 
beam pattern, and a Gaussian beam can be used as a representative case.
In Equation~(\ref{eq:gaussian_beam}), $\theta_0$ is the FWHM of the beam
and relates to the standard deviation of the Gaussian, $\sigma$, through
$\theta_0 = 2 \sqrt{2 \ln 2} \sigma$ .
We assume that the beam area on the sky is small enough that 
the solid angle can be integrated on a plane.
Then Condon (1974) derived the effective beam size as
\begin{eqnarray}\label{eq:omega_eff_gaussian}
  \Omega_{\rm eff} 
  &=& \int_{0}^{2\pi}\int_{0}^{\infty}
  e^{-(4\ln 2)(\gamma - 1)\left(\frac{\theta}{\theta_0}\right)^2}
  \theta d \theta d \phi 
  \nonumber \\
  &=& \pi \theta_0^2 \int_{0}^{\infty}
  e^{-(4\ln 2)(\gamma - 1)\psi} d \psi \nonumber \\
  &=& \dfrac{\pi \theta_0^2}{(4\ln 2)(\gamma - 1)}\; ,
\end{eqnarray}
$( 1 < \gamma < 3)$.
In the third step we set $\psi = (\theta/\theta_0)^2$.
We obtain the confusion limit flux to a cutoff deflection $x_{\rm c}$:
\begin{eqnarray}
  \sigma(x_{\rm c})^2 &=& \int_{0}^{x_{\rm c}} x^2 R(x) d x \nonumber \\
  &=& \int_{0}^{x_{\rm c}} \alpha \Omega_{\rm eff} x^{2 - \gamma} d x 
  = \alpha \Omega_{\rm eff} \int_{0}^{x_{\rm c}} x^{2 - \gamma} d x 
  \nonumber \\
  &=& \left( \dfrac{\alpha \Omega_{\rm eff}}{3 - \gamma} \right) 
  x_{\rm c}^{3 - \gamma} \; ,
\end{eqnarray}
and thus
\begin{eqnarray}
  \sigma(x_{\rm c}) =
  \left( \dfrac{\alpha \Omega_{\rm eff}}{3 - \gamma} \right)^\frac{1}{2} 
  x_{\rm c}^{\frac{3 - \gamma}{2}}.
\end{eqnarray}
As usually defined, if we set $x_{\rm c} = q \sigma$, we have 
\begin{eqnarray}
  \sigma =
  \left( \dfrac{\alpha \Omega_{\rm eff}}{3 - \gamma} \right)^\frac{1}{2} 
  q^{\frac{3 - \gamma}{2}} \sigma^{\frac{3 - \gamma}{2}}\; . \nonumber
\end{eqnarray}
Thus we obtain the important expression for the case of a power-law 
number count,
\begin{eqnarray}\label{eq:confusion_for_power}
  \sigma =
  \left( \dfrac{q^{3 - \gamma}}{3 - \gamma} \right)^\frac{1}{\gamma -1} 
  (\alpha \Omega_{\rm eff} )^\frac{1}{\gamma - 1} \; . 
\end{eqnarray}
For the Gaussian beam pattern, eqs.~(\ref{eq:confusion_for_power}) and
(\ref{eq:omega_eff_gaussian}) lead to 
\begin{eqnarray}\label{eq:confusion_for_power_gauss}
  \sigma &=& 
  \left( \dfrac{q^{3 - \gamma}}{3 - \gamma} \right)^\frac{1}{\gamma -1} 
  \left( \dfrac{\pi\theta_0^2\alpha}{(4 \ln 2)(\gamma - 1)} 
  \right)^\frac{1}{\gamma - 1} \; . 
\end{eqnarray}
This formula is used very frequently, probably because it is expressed in 
a simple analytic function.
However, we should note that this derivation includes a mathematical flaw
because $\theta$ is integrated from 0 to $\infty$, even though $\theta$ must 
be small.
Fortunately, the above integration converges, but when we generalize the form 
of the number counts, it diverges, as we discuss below.
Therefore we should set a certain cutoff in the integration in the real 
calculation.
Thus, the confusion limit given by 
Equation~(\ref{eq:confusion_for_power_gauss}) is too pessimistic.

We now turn to formulate the confusion limit for the general number counts 
(Franceschini et al.\ 1989).
Equation~(\ref{eq:diff_nc_beam}) leads to
\begin{eqnarray}
  \sigma^2 &=& \int_{0}^{x_{\rm c}} x^2 R(x)\, d x = 
  \int_{0}^{x_{\rm c}} x^2 \left[ 
    \int_{\Omega_{\rm beam}} 
    n \left( \dfrac{x}{h(\theta)} \right) 
    \dfrac{d \Omega}{h(\theta)} 
  \right]d x \nonumber \\
  &\simeq& \int_{0}^{x_{\rm c}} x^2 \left[ 
    \pi \theta_0^2\int_{0}^{\psi_{\rm c}} 
    n \left( \dfrac{x}{e^{-(4\ln 2) \psi}} \right) 
    \dfrac{d \psi}{e^{-(4\ln 2) \psi}} 
  \right]d x \nonumber \\
  &=& \dfrac{\pi \theta_0^2}{4\ln 2} \int_{0}^{x_{\rm c}} x^2 \left[ 
    \int_{1}^{\eta_{\rm c}} n ( \eta x ) d \eta
  \right]d x \equiv \dfrac{\pi \theta_0^2}{4 \ln 2} I(x_{\rm c})\; ,
\end{eqnarray}
where $\psi_{\rm c}$ and $\eta_{\rm c}$ are the upper cutoff, which 
corresponds to the beam area and is set to obtain a numerically reasonable 
result.
The result strongly depends on the value of the cutoff.
Just as in the above discussion, we set $x_{\rm c} = q\sigma$, so
\begin{eqnarray}
  \sigma = \sqrt{\dfrac{\pi I (q\sigma)}{4 \ln 2}}\, \theta_0\; .
\end{eqnarray}
Thus, we obtain a general relation between the beam size and the 
confusion limit as 
\begin{eqnarray}
  \theta_0 = \sqrt{\dfrac{4 \ln 2}{\pi I (q\sigma)}}\, \sigma \; .
\end{eqnarray}

\subsection{Required Survey Area}\label{subsec:area}

Next we formulate the calculation for the necessary sky area of the survey 
for the studies of galaxy evolution.  
In order to estimate the effects of galaxy evolution from survey data, 
a significant sky area should be scanned to suppress the variation in the 
galaxy surface number density.
Since galaxies are clustered on the sky, the nominal error bar estimated from 
the assumption of Poisson statistics is an underestimate, and we must 
take the galaxy angular correlation function into account.
Barcons (1992) considered the effect of clustering in his fluctuation analysis.
Despite its importance, this problem is often overlooked in the literature
on deep surveys.
At the faintest fluxes, this additional effect is gradually diluted by 
the projection along the line of sight.
We formulate the effect in the following.

Consider a survey area $\Omega$ and divide it into small cells 
$\{ \Delta\Omega_i \}$ so that the number of galaxies in the cell
$\{ \Delta\Omega_i \}$, ${\cal N}_i = 0$ or 1.
We set the mean galaxy surface number density ${\cal N}$.
Then we have a mean number in a solid-angle cell $\Delta\Omega_i$, 
${\cal N}_i$, 
\begin{eqnarray}
  \langle {\cal N}_i \rangle = {\cal N} \Delta\Omega_i\;.
\end{eqnarray}
By definition, we have $\langle {\cal N}_i \rangle = 
\langle {\cal N}_i^2 \rangle =   \langle {\cal N}_i^3 \rangle = \cdots$.
We observe
\begin{eqnarray}
  \langle {\cal N}_i {\cal N}_j \rangle &=&
  {\cal N}^2 (1 + w(\theta_{ij}))\,\Delta\Omega_i\Delta\Omega_j\\
  \langle ({\cal N}_i - \langle {\cal N}_i \rangle)
  ({\cal N}_j - \langle {\cal N}_j \rangle) \rangle &=&
    \langle {\cal N}_i {\cal N}_j \rangle - 
    \langle {\cal N}_i \rangle \langle {\cal N}_j \rangle \nonumber \\
    &=& {\cal N}^2 w(\theta_{ij})\,\Delta\Omega_i\Delta\Omega_j \; ,
\end{eqnarray}
where $w(\theta)$ is the angular two-point correlation function of
galaxies.
Next we consider the number of galaxies $N$ in the survey area $\Omega$.
We have
\begin{eqnarray}
  \langle N \rangle &=&
  \sum_{\Delta\Omega_i \subset \Omega} \langle {\cal N}_i \rangle = 
  \int_\Omega {\cal N} {\rm d} \Omega = {\cal N}\Omega\;,  \\
  \langle N^2 \rangle &=&
  \langle \sum_{i} {\cal N}_i \sum_{j} {\cal N}_j \rangle 
  \nonumber \\
  &=& \int_\Omega {\cal N} {\rm d} \Omega + 
  \int \int_\Omega {\cal N}^2 (1 + w(\theta_{12}))\,{\rm d} 
  \Omega_1 {\rm d} \Omega_2 \nonumber \\
  &=& {\cal N}\Omega + {\cal N}^2\Omega^2 + 
  {\cal N}^2 \int \int_\Omega w(\theta_{12})\,{\rm d} 
  \Omega_1 {\rm d} \Omega_2\;.
\end{eqnarray}
and thus 
\begin{eqnarray}
  \langle (N - \langle N \rangle)^2 \rangle =
  \langle N^2 \rangle - \langle N \rangle^2
  = {\cal N}\Omega + 
  {\cal N}^2 \int \int_\Omega w(\theta_{12})\,{\rm d} 
  \Omega_1 {\rm d} \Omega_2\;.
\end{eqnarray}
If we assume that we take sufficiently wide area whose dimension is 
much larger than the coherent scale of angular clustering, we can approximate 
the above expression as
\begin{eqnarray}
  \langle (N - \langle N \rangle)^2 \rangle &\simeq&
  {\cal N}\Omega + {\cal N}^2 \Omega \int_\Omega w(\theta)\,{\rm d} \Omega
   \nonumber \\
   &=& {\cal N} \Omega \left( 1 + {\cal N} \int_\Omega w(\theta)\,{\rm d} 
     \Omega \right)\;.
\end{eqnarray}
Here we evaluate the `signal-to-noise ratio' of the number counts 
${\sf S/N}$:
\begin{eqnarray}\label{eq:sn}
  {\sf S/N} &\equiv& \frac{\langle N \rangle}{
    \sqrt{\langle (N - \langle N \rangle)^2 \rangle}} \simeq 
  \frac{{\cal N}\Omega}{\sqrt{{\cal N}\Omega
      \left( 1 + {\cal N} \int_\Omega w(\theta)\,{\rm d} \Omega \right)}} 
  \nonumber \\
  &\simeq& \sqrt{\frac{\Omega}{
      \int_\Omega w(\theta)\,{\rm d} \Omega }}\;.
\end{eqnarray}
The approximation used in the third step can be adopted when ${\cal N}\Omega$
is sufficiently large.
We see that the ${\sf S/N}$ depends on the angular correlation strength and
the solid angle of the survey and is almost independent of the surface 
density of the sources ${\cal N}$.
This fact shows that, in order to determine galaxy evolution from 
number counts, large-area surveys are required.
The angular correlation function of the {\sl IRAS}~galaxies is $w(\theta) =
(\theta/\theta_0)^{-0.66}$ ($\theta_0 = 0.\hspace{-3pt}{}^\circ11$: 
Lahav, Nemiroff, \& Piran 1990).
We assume that the submillimeter sources and the faint radio sources have 
the same clustering properties as \iras~galaxies.
But even if we observe the same species of galaxies as the \iras~galaxies, 
an IR galaxy appears to be bright at FIR but faint at submillimeter;
therefore, we should assume their SEDs and convert the correlation function.
For the relatively bright radio sources, a steeper clustering power-law 
exponent $w(\theta) \propto \theta^{-(1.1 \mbox{--} 1.4)}$ has been reported 
by Cress et al.~(1996).
The angular correlation of galaxies out to a fainter flux limit is
obtained by using the scaling relation of $w(\theta)$ with a detection limit
through the relativistic Limber's equation (e.g., Peebles 1980).

Using this formula, in Section~\ref{subsec:int_z} we estimate the minimum 
survey area required for a sufficient ${\sf S/N}$.

\section{RESULTS AND DISCUSSIONS}\label{sec:result}

\subsection{The Galaxy Number Count Predictions Expected in the ALMA/LMSA}

\begin{figure*}[t]
\centering\includegraphics[width=7cm,angle=90]{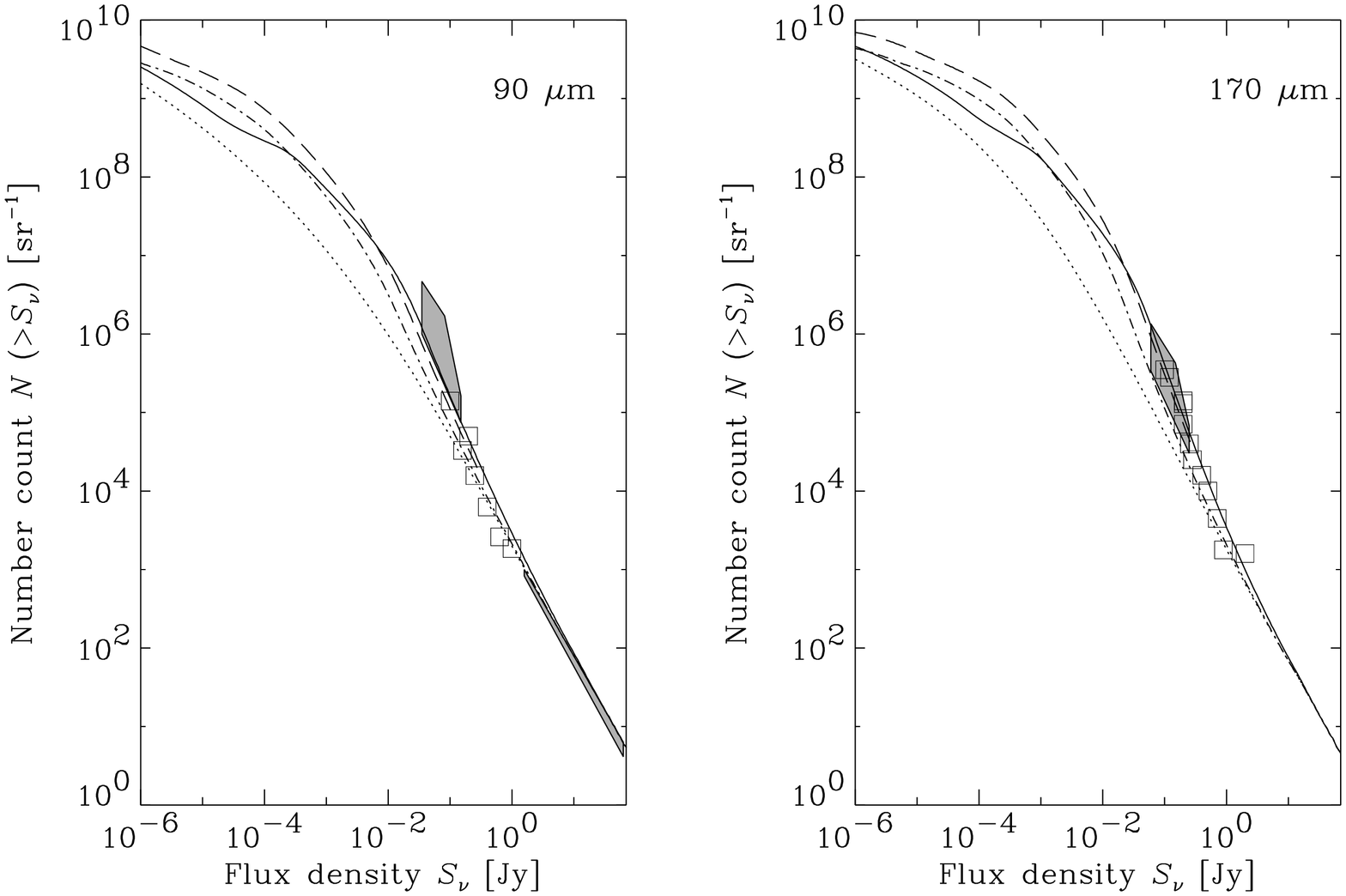}
\centering\includegraphics[width=7cm,angle=90]{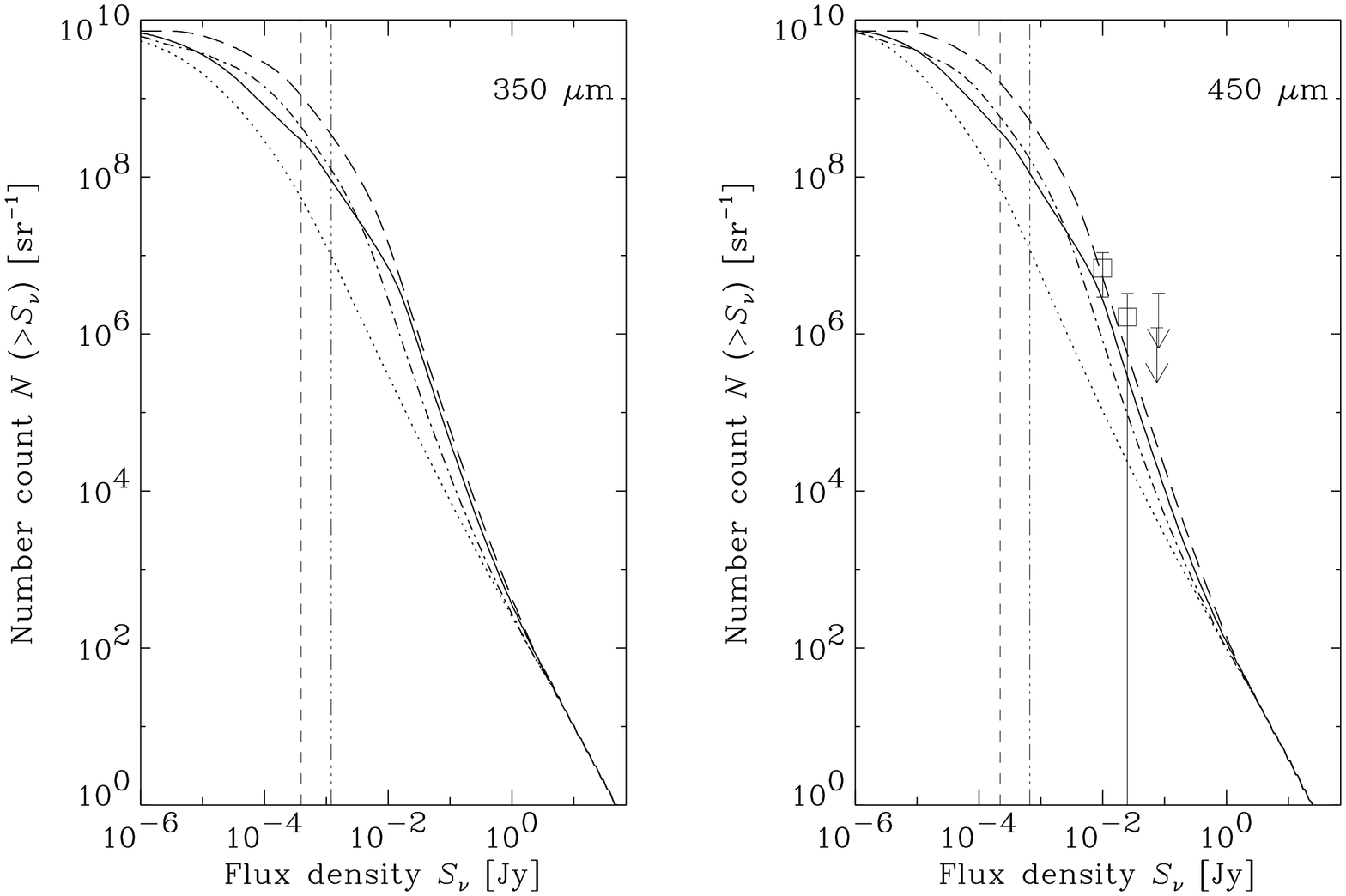}
\figurenum{3}
\figcaption[fig3a.ps,fig3b.ps,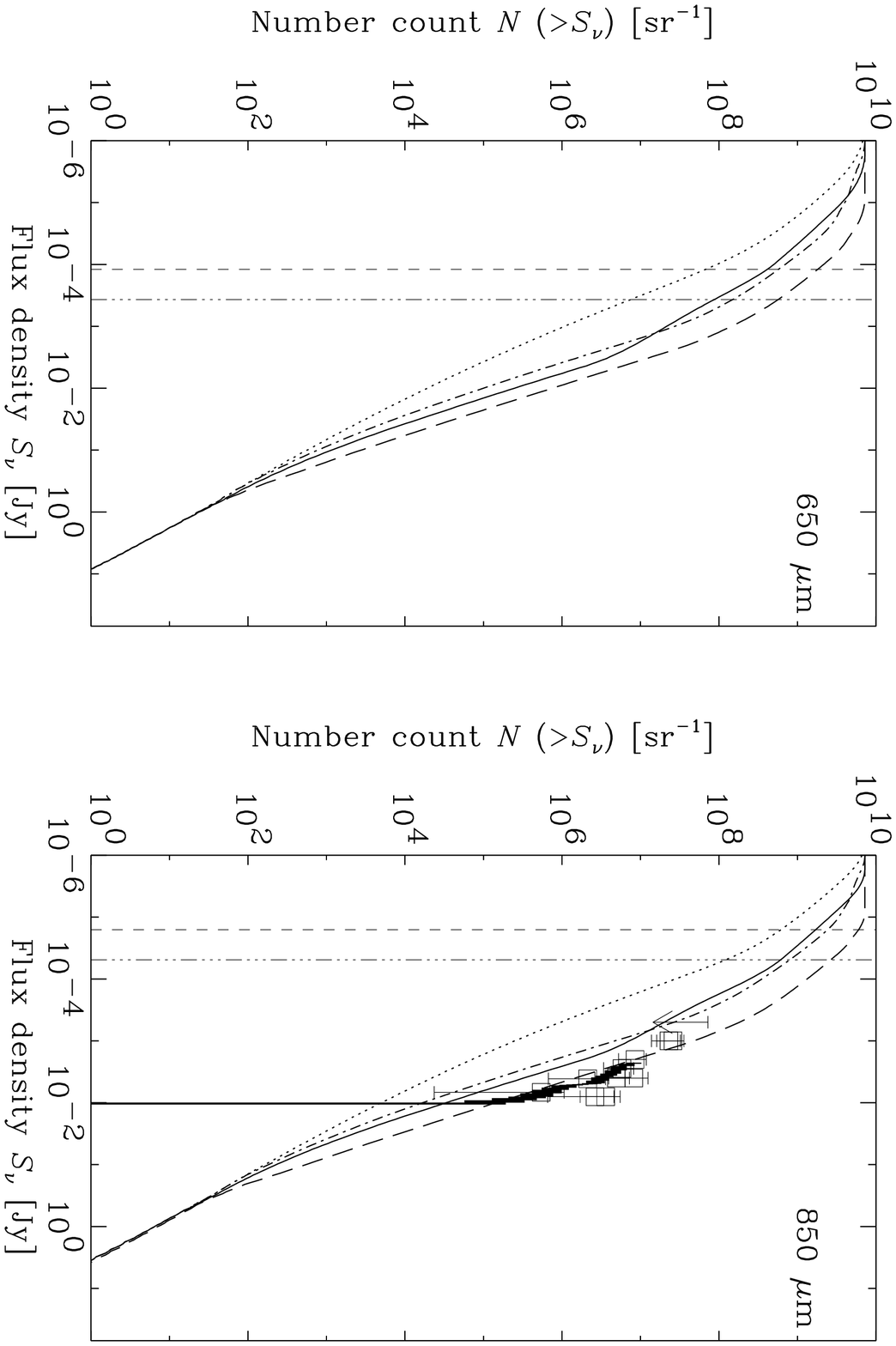,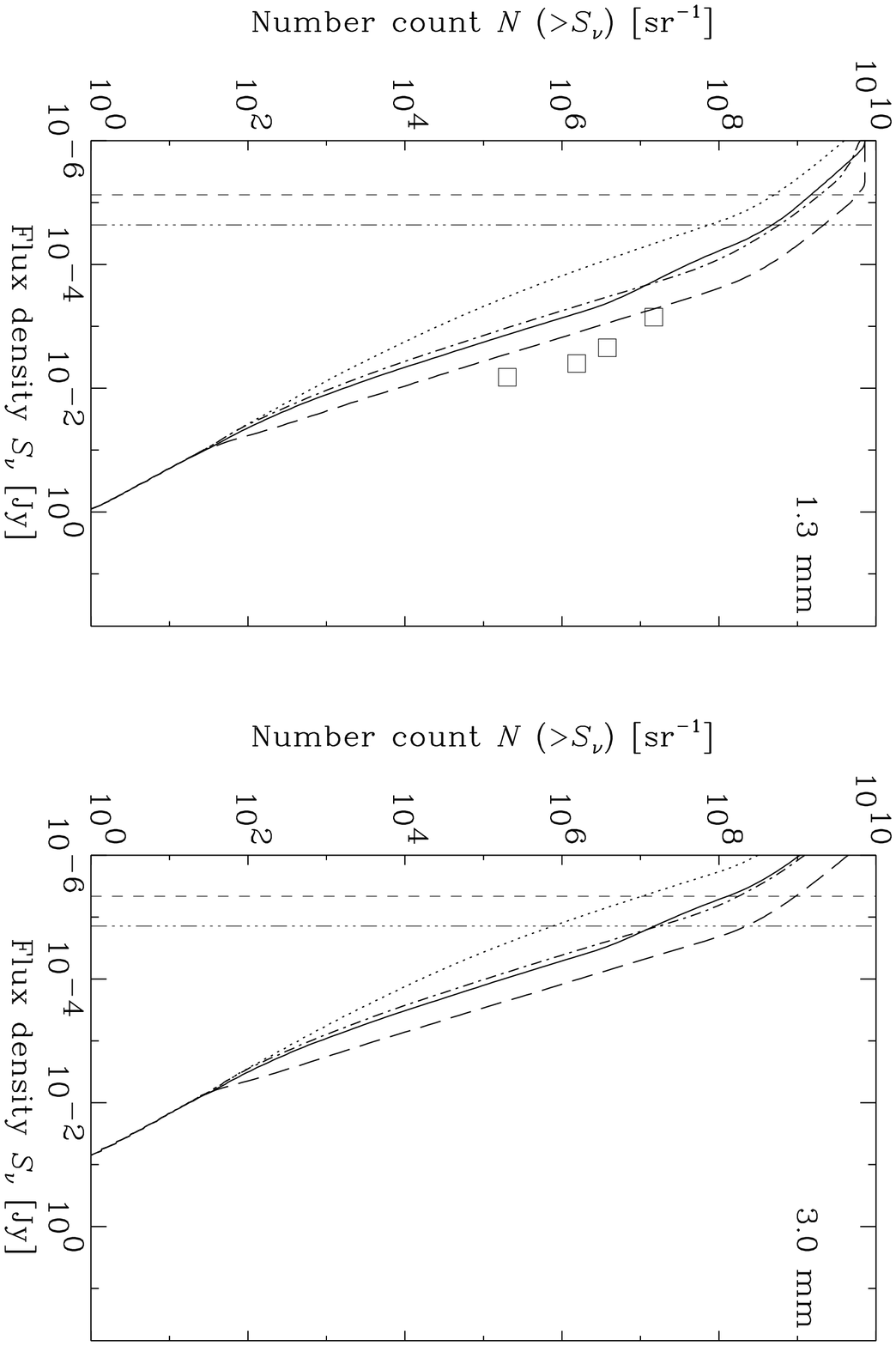]{
  The galaxy number count predictions from the infrared to the radio 
  wavelengths.
  The dotted curves represent the number counts of galaxies without evolution.
  The dot-dashed, solid, and long-dashed curves describe the number counts
  with Evolution 1, 2, and 3, respectively.
  In the infrared (90 and 170~$\mu$m), we plot the galaxy counts from 
  Stickel et~al.~(1998), 
  Kawara et~al. (1998), Puget et~al. (1999), Dole et~al. (2000), 
  Juvela, Mattila, \& Lemke (2000), Matsuhara et~al. (2000), and 
  Efstathiou et al. (2000).
  We also show the {\sl IRAS}~100 $\mu$m galaxy counts by a hatched thin area 
  at brightest fluxes.
  In the submillimeter (350, 450, 650, and 850~$\mu$m), the data are taken 
  from Smail et al. (1997), Barger et al. (1998), Barger, Cowie, \& Sanders 
  (1999), Hughes et al. (1998), Eales et al. (1999), Holland et al. (1999), 
  and Blain et al. (1999, 2000).
  At 1.3~mm, the preliminary result of MAMBO (Max-Planck Millimeter 
  Bolometer array) reported by Bertoldi (2000) is plotted.
  The vertical dot-dot-dot-dashed lines represent the $5\sigma$-detection 
  limits of the LMSA, and vertical dashed lines shows the 
  $5\sigma$-detection limits of the ALMA.
}\label{fig:nc}
\end{figure*}

First we show the galaxy number counts [$\,\mbox{sr}^{-1}$] from 
$90 \;\mu$m to 3~mm expected by the ALMA/LMSA in Figure~\ref{fig:nc}.
The dotted curves describes the no-evolution predictions at each wavelength.
The other curves depict the expected number counts with the three 
evolutionary histories discussed in Section~\ref{subsec:NC}, Evolution 1, 2, 
and 3.
All of these evolutions give a good fit to the FIR data and we cannot 
distinguish between each other with the present quality of the observed 
data (T01).
Some observed source counts are also shown in Figure~\ref{fig:nc}.
In the IR, we plot the galaxy counts from Stickel et~al.~(1998), 
Kawara et~al.~(1998), Puget et~al.~(1999), and Dole et~al.~(2000), Juvela, 
Mattila, Lemke (2000), Matsuhara et~al.~(2000), and Efstathiou et al.\ (2000).
We also show the {\sl IRAS}~100 $\mu$m galaxy counts by a hatched thin area 
at brightest flux.
In the submillimeter, the data are taken from Smail et al. (1997), 
Barger et al. (1998), Barger, Cowie, \& Sanders (1999), Hughes et al. (1998), 
Eales et al. (1999), Holland et al. (1999), and Blain et al. (1999, 2000).
At 1.3~mm, we show the preliminary result of MAMBO (Max-Planck Millimeter 
Bolometer array) reported by Bertoldi (2000).

\begin{figure*}[t]
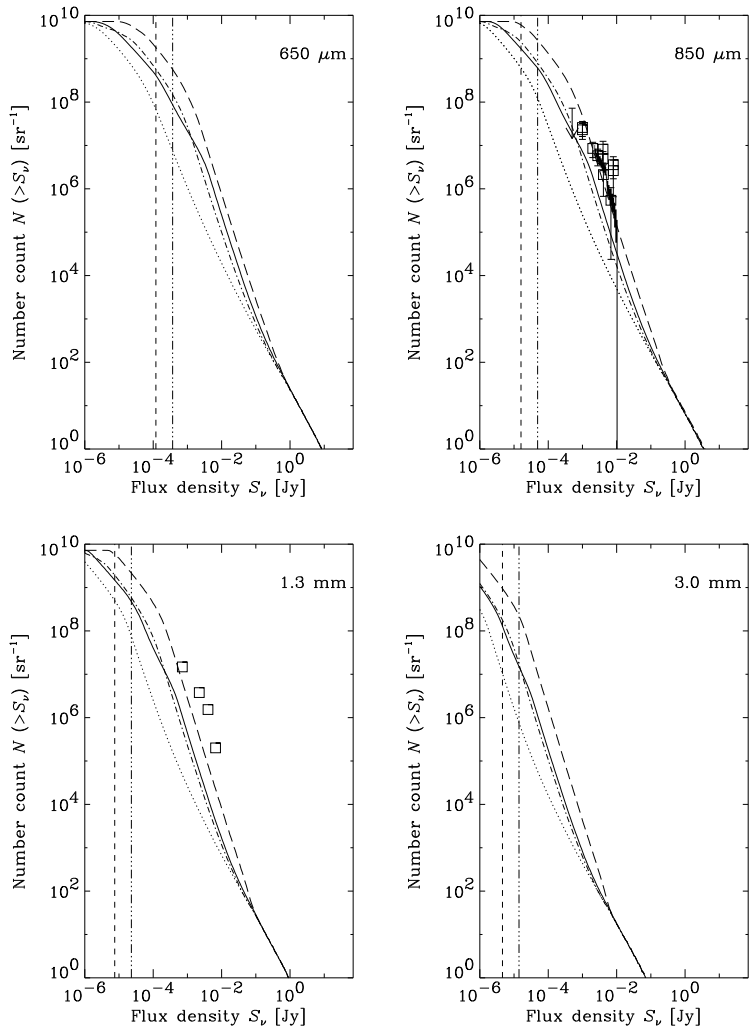

\figurenum{3}
\centering\includegraphics[width=7cm,angle=90]{fig3c.ps}
\centering\includegraphics[width=7cm,angle=90]{fig3d.ps}
\figcaption[fig3c.ps,fig3d.ps]{Continued.}
\end{figure*}

\subsection{The Source Confusion Limits}

\begin{figure*}[t]
\centering\includegraphics[width=8cm]{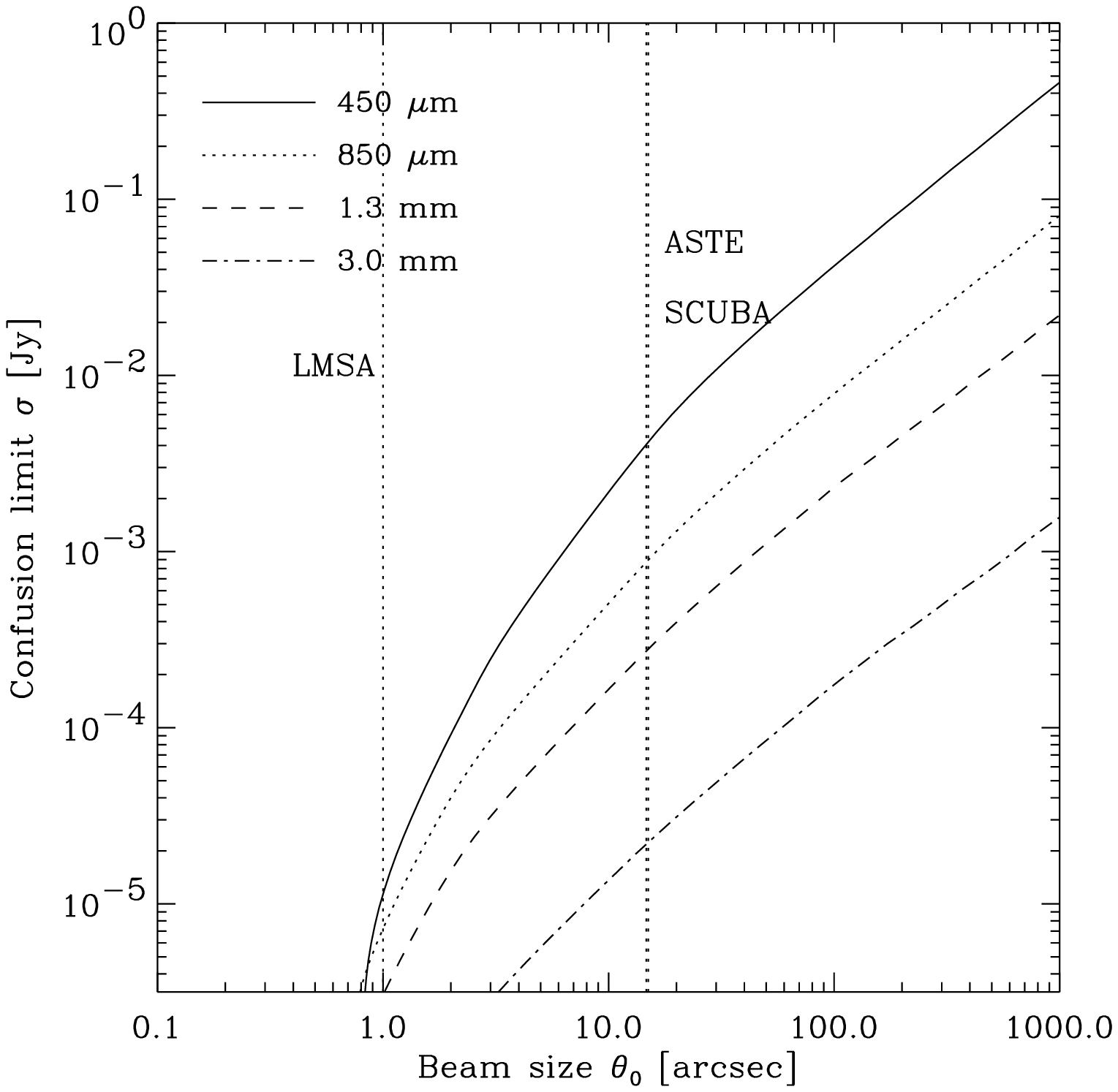}
\figurenum{4}
\figcaption[fig4.ps]{
  The $5\sigma$-source confusion limits as a function of the 
  beam size, calculated from the Evolution~2 in Figure~\ref{fig:evlp}.
  The solid, dotted, dashed, and dot-dashed curves represent the 
  confusion limit at $450\;\mu$m, $850\;\mu$m, 1.3 mm, and 3.0 mm, 
  respectively.
  The vertical dotted lines represent the angular resolutions
  of ASTE, SCUBA, and LMSA in the submillimeter wavelengths.
}\label{fig:confusion}
\end{figure*}

We calculated the source confusion limits at the submillimeter and radio 
wavelengths based on Evolution~2.
The results are shown in Figure~\ref{fig:confusion}.
The angular resolution of the ALMA will be better than $0_{\cdot}^{''}01$,
i.e., there will never be a confusion problem in the ALMA observation, 
and we do not show it in Figure~\ref{fig:confusion}.
When the slope of the number counts is flat, the source confusion limit 
will be much improved if better angular resolution can be achieved.
On the other hand, in the case that the slope of the number counts is very 
steep, the source confusion limit will not be drastically improved even with
a much better angular resolution.
Since the slope of the submillimeter counts is steep in the brighter flux 
regime and gets flatter towards the fainter fluxes, the confusion limit 
becomes lower quickly if the beam size (FWHM) becomes small.

In the submillimeter regime, we see that SCUBA ($\mbox{beam size} = 14.5\;
\mbox{arcsec}$) deep surveys are nearly confusion-limited.
Recently Hogg (2000) pointed out that if the number count slope is steep, 
i.e., $d \log N/d \log S \lnyoro -1.5$, the information of faint sources
can be completely destroyed by the confusion effect.
Eales et al. (2000) also performed a Monte Carlo simulation and concluded
that their estimated fluxes of the submillimeter SCUBA sources are actually 
boosted upwards significantly.
The median boost factor is 1.44, with a large scatter;
in the worst case the bias reaches an order of magnitude.
They also showed from their simulations that 19 \% of their SCUBA sources 
have positional errors greater than 6 arcsec.
Therefore, we should be quite cautious of the fluxes or positions of
the faintest submillimeter sources.
Considering this point, we can obtain the flux safely for sources brighter 
than $\sim 10$~mJy.
In the next subsection, we see that a reliable survey at brighter 
fluxes is crucially important for the study of galaxy evolution.
The large-area submillimeter survey of sources brighter than 10~mJy 
by the ASTE is suitable for this purpose.

On the other hand, the high angular resolution of the ALMA/LMSA results in 
a very low confusion limit of $< 1\;\mu$Jy.
Thus, we can estimate the flux and position of such faint sources
very precisely.
Since the slope of the number counts of millimeter sources becomes flatter at 
such faint fluxes, improving the angular resolution provides a drastically 
better confusion limit.

\subsection{Submillimeter Bright Source Counts and Galaxy Evolution at 
$z = 1 \sim 2$}\label{subsec:int_z}

\begin{figure*}[t]
\centering\includegraphics[width=10cm]{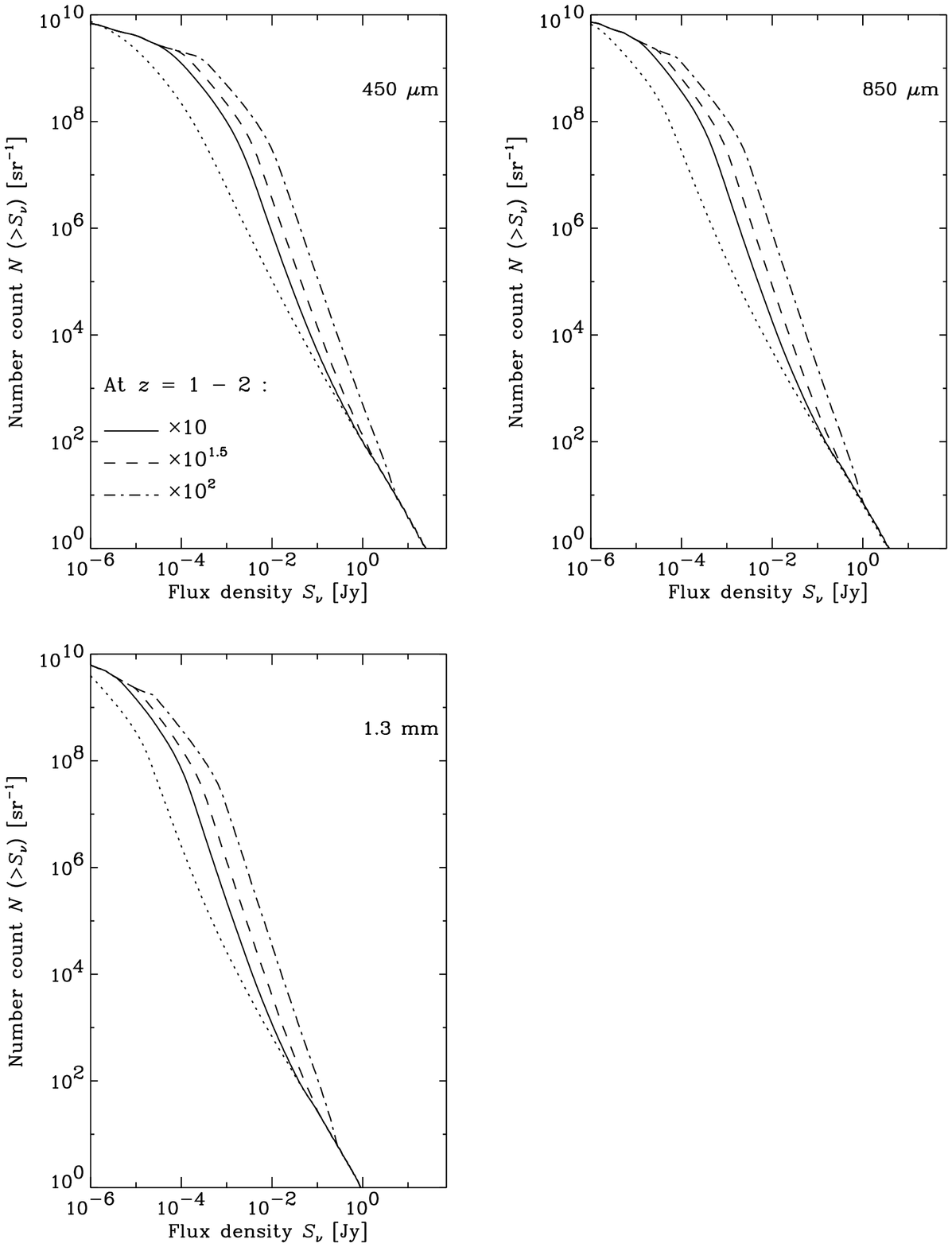}
\figurenum{5}
\figcaption[fig5.ps]{
  The resulting number counts produced by the evolutionary factor
  in the top panel of Figure~\ref{fig:evlp}.
  The dotted line shows the no-evolution result.
  The solid curve represents the galaxy number counts with Evolution~1.
  The dashed and dot-dashed lines represent the number counts with the
  modified evolutions so that the evolutionary factor at $z = 1 - 2$ 
  is $10^{1.5}$ and $10^{2.0}$, respectively.
  The effect of the evolution at $z = 1 - 2$ is clearly seen in 
  the bright end of the number count.
}\label{fig:nc_level}
\end{figure*}

In order to derive information on evolution at $z = 1 \sim 2$ 
from submillimeter source counts, we modify the evolutionary factor
and examine how the number counts vary with the modification.
We show the evolutionary factors used here in the top panel of 
Figure~\ref{fig:evlp}.
The thick solid step function is Evolution~1 itself.
We first modify Evolution~1 so that the evolutionary factor at $z = 1 - 2$ 
is $10^{1.5}$. 
Secondly we set the evolutionary factor at $z = 1 - 2$ to be $10^{2.0}$.
The resulting number counts are shown in Figure~\ref{fig:nc_level}.
The effect of the different evolutions at $z = 1 - 2$ is clearly seen in 
the bright end of the number counts in Figure~\ref{fig:nc_level}.
However an important problem arises here.
For measuring the number counts in such a bright flux regime, a very 
large-area sky survey is required.
Since the number density of the sources is so small and the statistical 
fluctuations are large, secure determination of the counts is difficult.

The necessary sky area is estimated by using the formulation presented 
in Section~\ref{subsec:area}.
The characteristic depth of a survey is defined by the survey flux limit, 
$S_{\rm lim}$.
As we mentioned in Section~\ref{subsec:area}, we here assume that the
submillimeter sources brighter than 100~mJy are clustered as strongly as the 
\iras~sources brighter than 0.7~Jy, which have the angular correlation 
function $w_0 (\theta) = (\theta/\theta_0)^{-0.66}$.

\begin{figure*}[t]
\centering\includegraphics[width=7cm,angle=90]{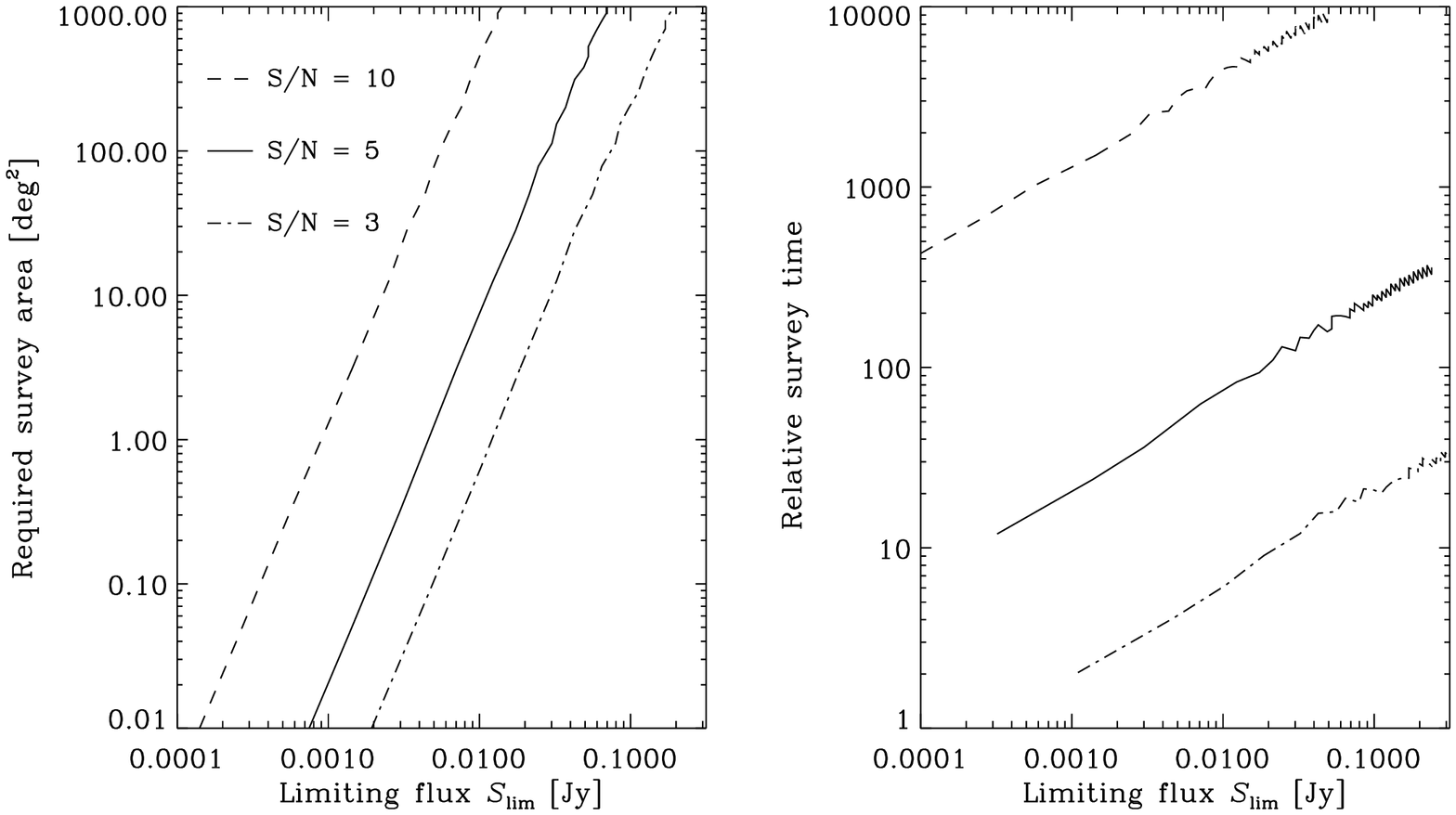}
\figurenum{6}
\figcaption[fig6.ps]{
  {\sl Left panel}: The required sky area to suppress the galaxy 
  clustering fluctuation as a function of limiting flux of a survey.
  {\sl Right panel}: Relative observing time to perform the corresponding 
  survey.
  The unit of the time is arbitrary.
}\label{fig:area}
\end{figure*}

We define the relative characteristic depth of a survey by $d_* = 
(0.1\;[\mbox{Jy}]/S_{\rm lim}\;[\mbox{Jy}])^{1/2}$.
Then, the angular clustering $w(\theta)$ is scaled as
\begin{eqnarray}
  w(\theta) = d_*^{-1} w_0 (d_*\theta)
\end{eqnarray}
where $w_0(\theta)$ is the angular correlation function of the survey
whose flux limit is 0.1~Jy.
Under the assumptions mentioned above, the clustering of the submillimeter 
sources is expressed as
\begin{eqnarray}\label{eq:relcor}
  w(\theta) = d_*^{-1.66} w_0 (\theta)\;.
\end{eqnarray}
The signal-to-noise ratio defined in Section~\ref{subsec:area} becomes
\begin{eqnarray}
  {\sf S/N} = \left( \dfrac{\Omega}{
      d_*^{-1.66} \int_\Omega w_0(\theta)d \Omega}\right)^{0.5}
  = \left( \dfrac{(S_{\rm lim}/0.1)^{-0.83}\Omega}{
      \int_\Omega w_0(\theta)d \Omega}\right)^{0.5}\;,\nonumber
\end{eqnarray}
and we obtain
\begin{eqnarray}\label{eq:approxslim}
  S_{\rm lim} = 0.1 \,\left( \dfrac{\Omega}{
      ({\sf S/N})^2 \int_\Omega w_0(\theta)d \Omega}\right)^{1/0.83}\;.
\end{eqnarray}
In the left panel of Figure~\ref{fig:area} we show the area--limiting flux 
relation derived directly from eqs.~(\ref{eq:sn}) and (\ref{eq:relcor}).
In fact, the equation is well approximated by 
\begin{eqnarray}
  \Omega (S_{\rm lim})\propto S_{\rm lim}^{2.5}\;.
\end{eqnarray}
The above approximation is valid as long as the fluctuation by 
clustering dominates the variance.
Since recent observations of the submillimeter number counts suggest 
strong galaxy evolution (Section~\ref{sec:introduction}), the number counts 
increase rapidly toward fainter fluxes.
Consequently, the Poisson fluctuation is small as compared with the clustering 
variance and the condition for the approximation is satisfied.
Thus, the relation presented in Figure~\ref{fig:area} is well represented
by Equation~(\ref{eq:approxslim}).

We can also estimate the relative observation time to perform such a survey
by multiplying $S_{\rm lim}^2$ by the obtained sky area $\Omega (S_{\rm lim})$.
It is straightforward to see that the relative observation time is roughly 
proportional to $\sqrt{S_{\rm lim}}$.
Therefore, the fainter the limiting flux, the shorter the required 
observation time.
This is presented in the right panel of Figure~\ref{fig:area}.
But actually, as we discussed above, we cannot make the flux limit too 
faint ($\gnyoro 10$~mJy) because of severe source confusion.

If we adopt ${\sf S/N} = 5$, the required survey area is $10\;{\rm deg}^2$
at 10~mJy, which is large but not unreasonable for a present-day submillimeter 
survey.
If ${\sf S/N} = 3$, then the survey area is $20\;{\rm deg}^2$ at 50~mJy.
Thus, the large-area survey by the ASTE is suitable for this purpose.
For the minimal survey of $1\;\mbox{deg}^2$, the virtual observation time 
is estimated to be approximately 200 hours by a single-pixel detector.
If the array-type detector is available, this survey will be
performed much more quickly and efficiently.

We note that these predictions depend on the assumed clustering strength.
When the survey is finished, we will have a large database of submillimeter 
sources with firm counterpart associations.
Using the resulting catalog, we will be able to estimate the 
angular clustering properties of the submillimeter sources.
If the variance of the detected source counts are different than what we 
assumed, then it provides important quantitative information on the 
clustering evolution of dusty galaxies.
We stress that this will be another important purpose of this survey.

\subsection{Faintest Source Counts and Galaxy Formation}\label{subsec:zmax}

We next adopt three other cases for the evolutionary factor $f(z)$ 
to examine the effect of the galaxy formation epoch on the number counts 
at submillimeter and radio wavelengths.
First we use the evolutionary histories based on Evolution~2, and vary 
the the cutoff redshifts as $z_{\rm form} = 2$, 3, 5, and 7.
We note that this cutoff redshift represents the epoch of the formation or 
appearance of the dusty galaxies, and does not necessarily imply the formation
of all species of galaxies.
In addition, we also study the case of submillimeter-based Evolution~3,
introduced in Section~\ref{subsec:NC}.
For the modified evolutionary histories based on the Evolution~3,
the redshift cutoff are also set to be $z_{\rm form} = 2$, 3, 5, and 7.

\begin{figure*}[t]
\centering\includegraphics[width=10cm]{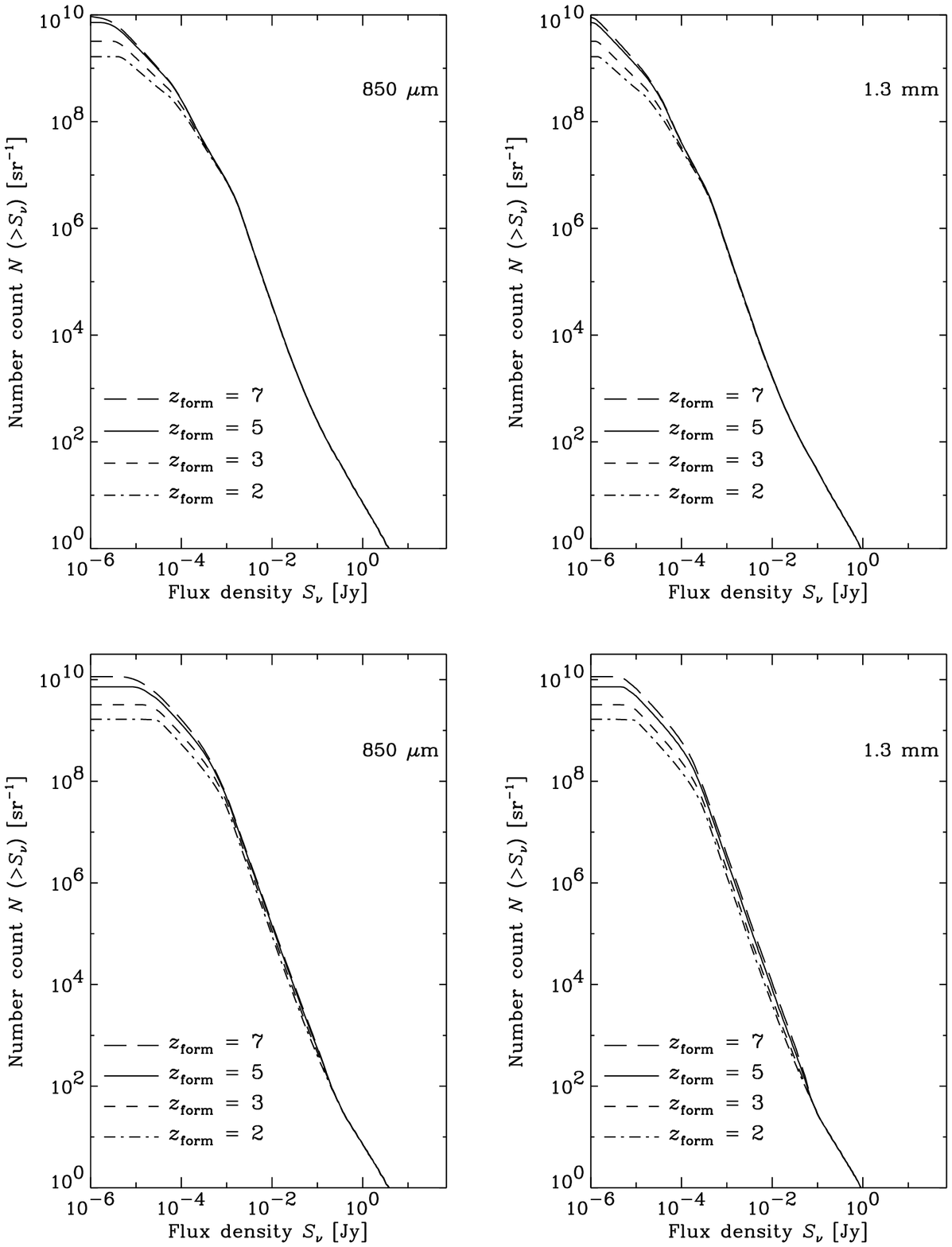}
\figurenum{7}
\figcaption[fig7.ps]{
  {\sl Top panels}: The galaxy number counts at 850~$\mu$m and 1.3~mm 
  with redshift cutoffs $z_{\rm form} = 2, 3, 5$, and 7 based on Evolution~2 
  in Figure~\ref{fig:evlp}.
  {\sl Bottom panels}: The same as the Top panels except that the number 
  counts are based on Evolution~3 in Figure~\ref{fig:evlp}.
  In both panels, the dot-dashed, dashed, solid, and long-dashed curves 
  represent the resulting number counts with redshift cutoffs 
  $z_{\rm form} = 2, 3, 5$, and 7, respectively.
  A very low detection limit is required to investigate the redshift 
  cutoff, i.e., the galaxy formation epoch.
}\label{fig:nc_zmax}
\end{figure*}

The resulting number counts with various redshift cutoffs are shown in 
Figure~\ref{fig:nc_zmax}.
The top two panels present the number counts at 850~$\mu$m and 1.3~mm
with Evolution 2.
The bottom tow panels show the number counts with Evolution 3.
The effect of the cutoff is clearly seen in the faintest end of the 
number counts in these two panels.
In the bottom panel of Figure~\ref{fig:nc_zmax}, the effect of redshift 
cutoff is more prominent than in the top panel.

In contrast to the case discussed in Section~\ref{subsec:int_z}, 
a very deep detection limit is required to investigate the redshift 
cutoff, i.e., the galaxy formation epoch.
The extremely deep survey by ALMA/LMSA is a unique opportunity 
for this purpose.
At such a faint flux level, the variance of the number count is diluted by the 
large source surface density and the source projection; we can, therefore,
distinguish between the predictions of different models presented in 
Figure~\ref{fig:nc_zmax}.

When we discuss such a deep flux limit, the dynamic range of a detector 
should be taken into account.
If a very bright source exists in the field of view, the faintest sources
near the detection limit cannot be detected in the field.
We evaluated the probability that the sources $10^3$ times brighter than 
the $5\sigma$-detection limit of the LMSA and ALMA exist in a field of view.
The field of view of the LMSA observation is 20-arcsec diameter circular
area ($7.4 \times 10^{-9}$ sr).
The number density of the sources $10^3$ times brighter than the $5\sigma$ 
limit at each waveband, $N$, is directly obtained from Figure~\ref{fig:nc}.
For example, $N$ of the ALMA at $850 \; \mu$m is 
$4 \times 10^4 \; \mbox{sr}^{-1}$.
Then we obtain the expectation value of the number of very bright
sources compared with the detection limit found in the field of view, 
$\mu = N \times (\mbox{field of view})$.
By using this value, we can treat the probability of finding $k$ bright 
sources in the field as a Poisson process
\begin{eqnarray}
  P_\mu (k) = \dfrac{e^{-\mu} \mu^k}{k!}\; .
\end{eqnarray}
Then the probability that more than one source exist in the field of view is 
\begin{eqnarray}
  1 - P_\mu (0) = 1 - e^{-\mu} \; .
\end{eqnarray}
Even in the worst case of the ALMA (when we use Evolution~3) the probability 
is $\sim 3 \times 10^{-4}$.
Thus, the probability that a very bright radio source lies in the field of 
view is so small that we do not have to worry about the dynamic range, 
and we can safely obtain the information of very distant faint sources.

\subsection{Redshift Estimation from the Dust Continuum}

Determination of source redshifts is an important but quite challenging issue 
in cosmological studies.
Information on source redshifts enables us to construct the LF as a function 
of redshift.
We here assume the LF shape to be unchanged during evolution, but
this is a strong assumption which should be tested with observational data.
Spectroscopic observations will be extremely difficult or almost impossible
on very distant objects, and alternative methods are necessary.
In the optical wavelength, photometric redshift technique has provided a
breakthrough for further studies of ultra high-redshift objects 
(e.g., Fern\'{a}ndez-Soto et al. 1999).

On the other hand, at longer wavelengths such as the far-IR (FIR) or 
submillimeter it is harder to estimate source redshifts because of 
the smooth nature of their SEDs.
Takeuchi et al. (1999) tried to estimate redshifts roughly from FIR 
photometry and found a practical possibility.
Carilli \& Yun (1999) proposed the radio-to-submillimeter spectral index 
as a redshift indicator for star-forming galaxies.
They used the correlation between the FIR flux from the thermal dust
and the radio flux from the synchrotron emission from the interstellar matter 
and supernova remnants, and utilized the spectral index between the two kinds 
of emission as a function of redshift.

Though the radio-to-submillimeter index method is now often used to estimate 
the redshift of submillimeter sources, the origin of the two components of 
radiation is substantially different.
If we can estimate the redshift of the sources from the ratio of two or
more flux densities which have the same physical origin, it will be 
undoubtedly the most desirable method.

We here develop such a method which is an extension of the attempt of 
Takeuchi et al.\ (1999).
We use FIR and submillimeter/millimeter flux densities for this purpose.
For the FIR database, we suppose the wideband photometric catalog of the 
ASTRO-F FIR all-sky survey.
At FIR to submillimeter wavelengths, continuum emission is dominated by
blackbody radiation from the big dust grains in equilibrium with the ambient
ultraviolet radiation field.
Thus, the flux densities in this wavelength range are radiated from the 
same emission mechanism.

\begin{figure*}[t]
\centering\includegraphics[width=14cm]{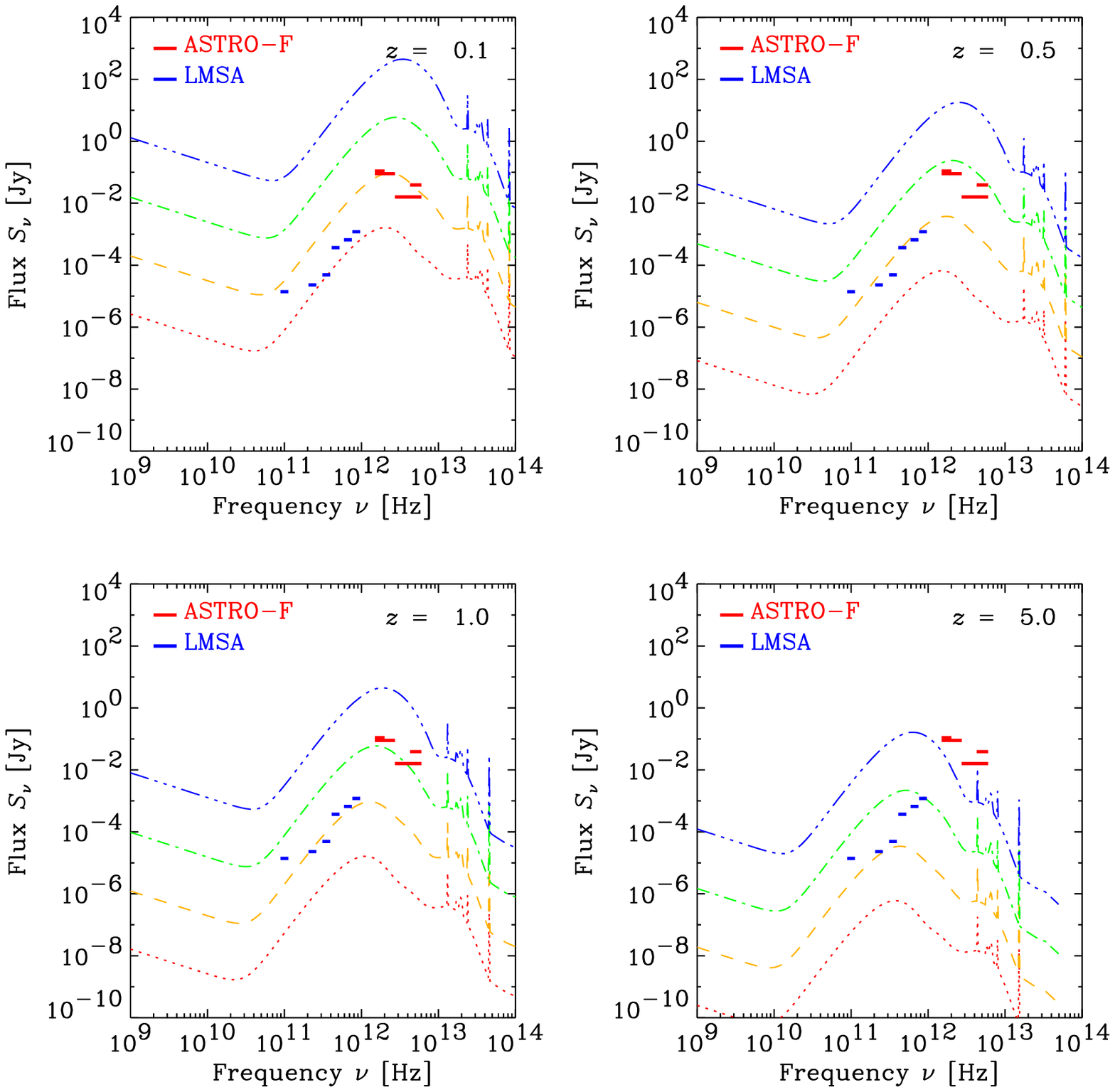}
\figurenum{8}
\figcaption[fig8.ps]{
  The redshifted SEDs of infrared galaxies with luminosities
  $10^{8}$, $10^{10}$, $10^{12}$, and  $10^{14}\,L_\odot$, at
  redshift $z = 0.1, 0.5, 1.0$, and 5.0
  (blue dot-dot-dot-dashed lines: the SED of galaxies with $10^{14}\;
  L_\odot$, green dot-dashed lines: $10^{12}\; L_\odot$, 
  yellow dashed lines: $10^{10}\; L_\odot$, and red dotted
  lines: $10^{8}\; L_\odot$).
  The $5\sigma$-detection limits of ASTRO-F and LMSA are also shown
  as thick horizontal short lines.
  The $5\sigma$ sensitivities of the LMSA at wavelengths 350 $\mu$m, 
  450 $\mu$m, 650 $\mu$m, 850 $\mu$m, 1.3 mm, and 3.0 mm (mean values in 
  winter season) are 1200, 660, 370, 49, 23, and 14 $\mu$Jy beam$^{-1}$, 
  respectively.
  The bandpass system of the far-IR (FIR) instrument, Far Infrared Surveyor 
  (FIS), consists of two narrow bands, N60 ($50 - 70\;\mu$m) and N170 
  ($150 - 200\;\mu$m), and two wide bands, WIDE-S ($50 - 110\;\mu$m) and 
  WIDE-L ($110 - 200\;\mu$m).
  The detection limits are estimated as 39~mJy and 110~mJy for N60 and N170, 
  and 16~mJy and 90~mJy for WIDE-S and WIDE-L, respectively.
}\label{fig:redshiftSED}
\end{figure*}

\begin{figure*}[t]
\centering\includegraphics[width=10cm]{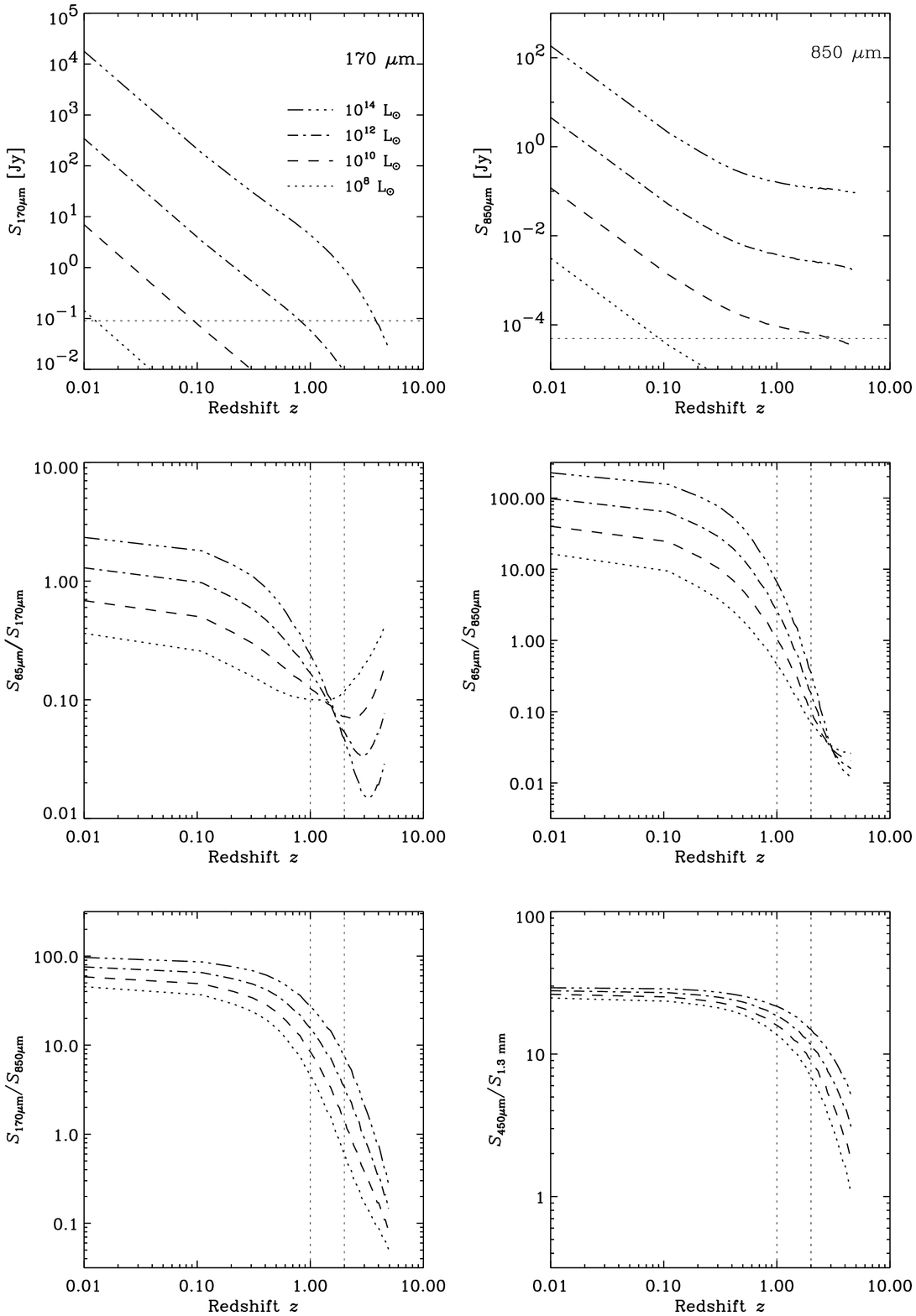}
\figurenum{9}
\figcaption[fig9.ps]{
  The flux--redshift and color--redshift diagrams of 
  the model infrared galaxies with luminosities same as 
  Figure~\ref{fig:redshiftSED}.
  Top two panels are the flux--redshift relations at $170\;\mu$m 
  and $850\;\mu$m.
  The horizontal dotted line of the top-left and top-right panels
  depicts the detection limit at each wavelength.
  Middle and bottom panels present the color--redshift relations.
  The vertical dotted lines represent the redshift $z = 1$ and 2.
}\label{fig:dust_zC}
\end{figure*}

First we show the redshifted SED of an infrared galaxy with luminosities
$10^{8}$, $10^{10}$, $10^{10}$, and  $10^{12}\,L_\odot$ at redshift 
$z = 0.1, 0.5, 1.0$, and 5.0 in Figure~\ref{fig:redshiftSED}.
The $5\sigma$-detection limits of ASTRO-F and LMSA are also shown.
The problem is that the peak of the blackbody radiation shifts not
only with redshift but also with dust temperature $T_{\rm dust}$.
Therefore, this method suffers from this degeneracy.
Is it useless as a redshift estimator?
In order to clarify this point, we present the detected flux densities and 
flux density ratios of star-forming galaxies as a function of redshift in 
Figure~\ref{fig:dust_zC}.
The top two panels are the flux--redshift relations at $170\;\mu$m 
and $850\;\mu$m.
The horizontal dotted line in the top-left and top-right panels
depicts the detection limit at each wavelength.
The middle and bottom panels present the color--redshift relations.
Submillimeter colors are a strong function of redshift, and we
expect that they work well as rough redshift indicators.

\begin{figure*}[t]
\centering\includegraphics[width=10cm]{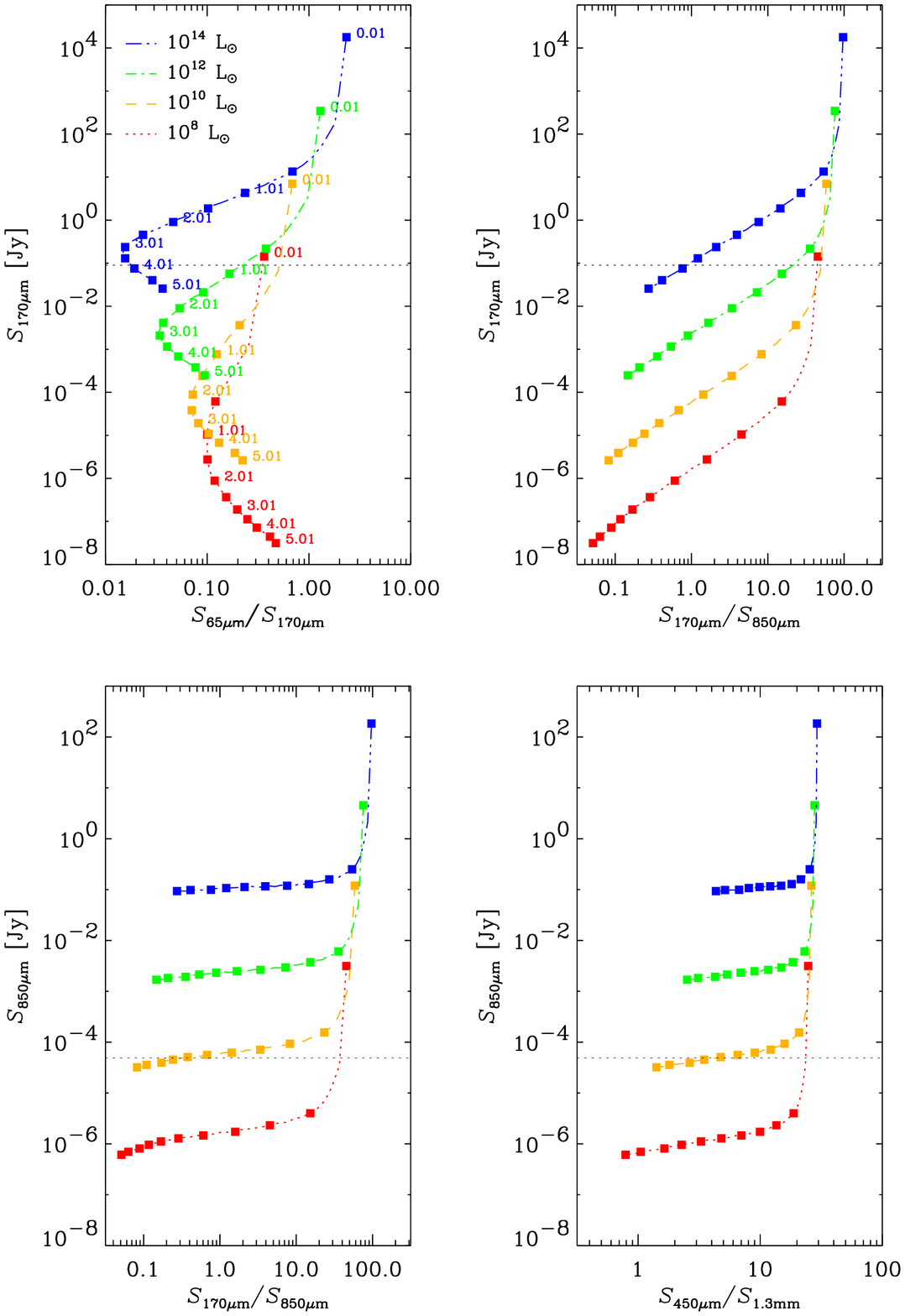}
\figurenum{10}
\figcaption[fig10.ps]{
  The color--flux diagrams of the redshifted model infrared galaxies
  at various luminosity and redshift.
}\label{fig:dust_CF}
\end{figure*}

But the ambiguity is of order unity and is too large for, e.g.,
estimating the shape evolution of the source LF.
Here we can use some additional information.
We use the empirical relation between FIR luminosities of galaxies and
the flux ratio $S_{60\mu{\rm m}}/S_{100\mu{\rm m}}$ 
(Equation~(\ref{eq:color_flux})) when we construct the SED model.
Therefore when we obtain the flux ratio of a source, we can estimate
the luminosity $L_\nu (\mbox{at }60\mu{\rm m})$.
Then we compare the estimated flux density
\begin{eqnarray}
  \hat{S}_{60 (1 + z)\mu{\rm m}} = 
  \frac{(1+z)\hat{L}_\nu(\mbox{at }60\;\mu{\rm m})}{4\pi \dl(z)^2}
\end{eqnarray}
with the detected flux density at the corresponding wavelength.
If the two flux densities are significantly different, we correct 
the assumed $T_{\rm dust}$ and repeat this trial.
Thus we can obtain the redshift estimation by an iterative procedure.
Galaxies do not degenerate in the color--color--flux three-dimensional
space (Figure~\ref{fig:dust_CF}), and we safely use this method.
We call this {\sl the dust-$z$ method}.

\begin{figure*}[t]
\centering\includegraphics[width=7cm,angle=90]{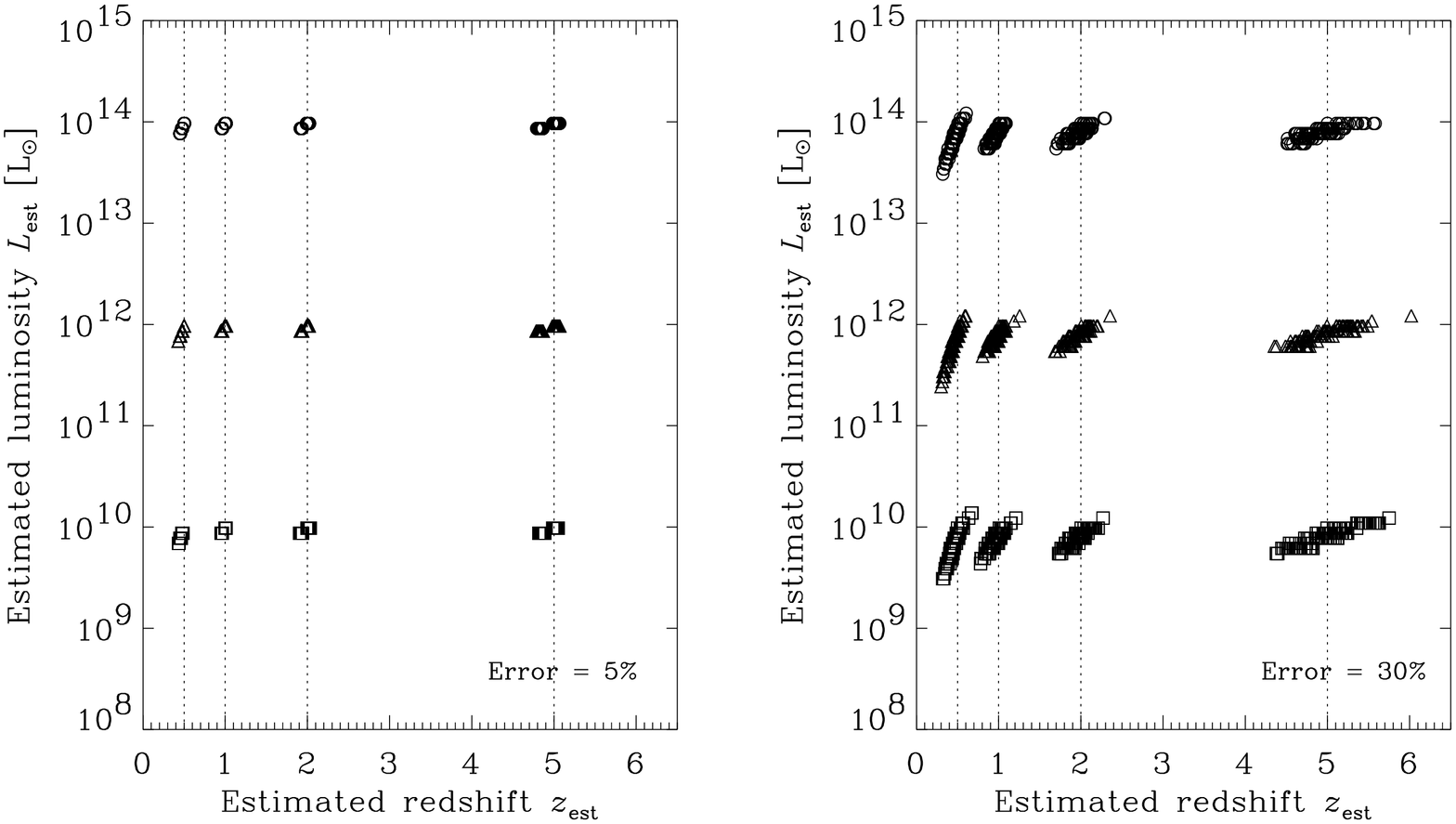}
\figurenum{11}
\figcaption[fig11.ps]{
  The Monte Carlo simulation of the dust-$z$ method.
  The open squares, triangles, and circles represent galaxies with input 
  luminosity $10^{10}, 10^{12}$, and $10^{14}\;L_\odot$, respectively.
  The vertical dotted lines depict redshifts of 0.5, 1.0, 2.0, and 5.0.
  Left panel shows that the added random errors are $\sim 5$~\%, and right
  panel presents the case that the errors are $\sim 30$~\%.
}\label{fig:dustz_sim}
\end{figure*}

In order to examine the accuracy of the dust-$z$ method, we performed a 
series of Monte Carlo simulations.
We assumed the following observational wavebands: 60, 90, 170, 450, 850, 
1300, and 3000~$\mu$m.
The errors in flux measurements are set to be 5~\% and 30~\%.
We input the assumed SEDs and added random errors to them, and calculate
their `observed' fluxes.
Then we performed the above algorighm, and estimated the redshifts and 
luminosities of simulated galaxies.
The result of the simulation is presented in Figure~\ref{fig:dustz_sim}.
We show the estimates for galaxies with input luminosity $10^{10}, 10^{12}$, 
and $10^{14}\;L_\odot$.
The input redshifts of galaxies are fixed to 0.5, 1.0, 2.0, and 5.0.
We examined two representative cases where the errors are $\sim 5$~\% and 
$\sim 30$~\%.
In these simulations, we calculated 100 realizations for each redshift and 
luminosity.
We conclude that if the error in each band is $\sim 5$~\% (i.e.\ $S/N \sim 
20$), then the redshift can be successfully estimated by the dust-$z$ method.
Even in case the error is $\sim 30$~\%, the uncertainty in the estimation
is comparable or even better than some other presently used indirect methods 
for redshift estimation (cf.\ Rengarajan \& Takeuchi 2001).

We also evaluate the uncertainty in the adopted SEDs.
Taking the error in Equation~(\ref{eq:color_flux}) into account, we assess
the error of $\sim 40$~\% in luminosity estimation.
The luminosity of galaxies in the FIR is well approximated by 
$L_{\rm FIR} \propto T_{\rm dust}^{4+\gamma}$, hence the error in temperature
estimation is at most $(1 \pm 0.4)^{1/4+\gamma} \sim 1 \pm 10$~\%.
The peak frequency, $\nu_{\rm peak}$, obeys the Wien displacement law:
$\nu_{\rm peak} = 5.88 \times 10^{10}\,T$.
Thus, the error associated with the uncertainty in SED templates is evaluated
as $\sim 10$~\%.

This method works effectively when we have many photometric bands or channels.
Therefore, instruments that have a large number of wavebands are
very useful for this purpose.
This method provides us not only the redshift information but also the dust
temperature of the submillimeter sources at the same time.
This will surely be a strong constraint on the star formation history of 
galaxies.

\subsection{$1.4$-GHz Radio Source Counts}

\begin{figure*}[t]
\centering\includegraphics[width=10cm]{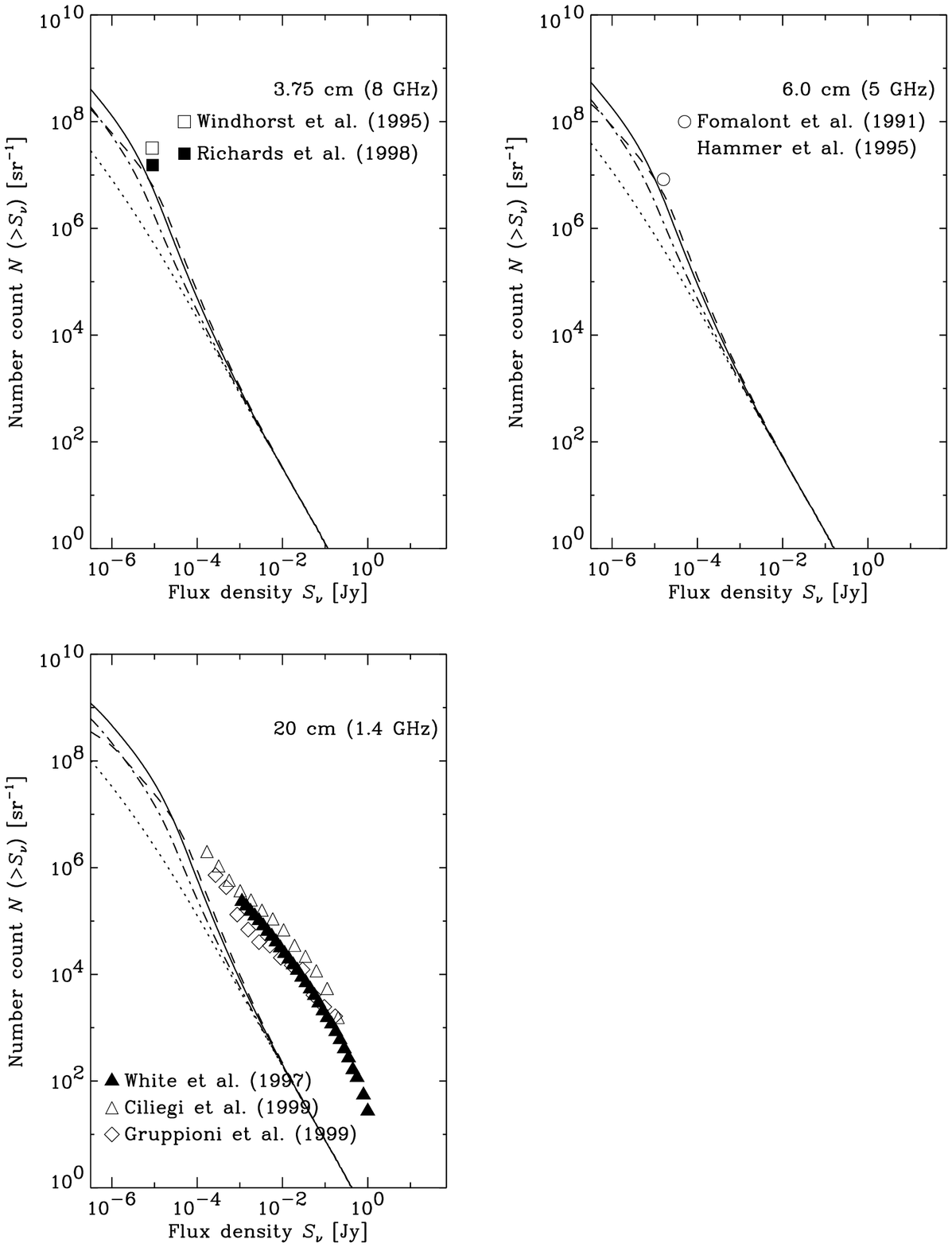}
\figurenum{12}
\figcaption[fig12.ps]{
  The comparison of the observed 8, 5, and 1.4~GHz source counts
  and our model counts.
  The same as Figure~\ref{fig:nc}, the dotted curves indicate the number 
  counts of galaxies without evolution, and the dot-dashed, solid, and 
  long-dashed curves represent the number counts with Evolution 1, 2, and 3, 
  respectively.  
  The observational data plotted are taken from Windhorst et al. (1995), 
  Richards et al.\ (1998) (8~GHz), Fomalont et al.\ (1991), 
  Hammer et al.\ (1995) (5~GHz), White et al.\ (1997), Ciliegi et al.\ (1999) 
  and Gruppioni et al.\ (1999) (1.4~GHz).
}\label{fig:nc_cm}
\end{figure*}

We show the comparison of the observed 8, 5, and 1.4~GHz source counts
and our model counts in Figure~\ref{fig:nc_cm} to examine the 
contribution of star forming galaxies to the faint radio galaxies.
The observational data are taken from Windhorst et al. (1995), 
Richards et al.\ (1998) (8~GHz), Fomalont et al.\ (1991), 
Hammer et al.\ (1995) (5~GHz), White et al.\ (1997), Ciliegi et al.\ (1999) 
and Gruppioni et al.\ (1999) (1.4~GHz).
At 8~GHz the observed source counts are slightly higher than our 
model prediction, but the discrepancy is not so significant because the 
survey areas are quite small in these studies.
The 5~GHz data are well reproduced by the present model.
Recent studies revealed that the faint radio sources with flux densities
$S_\nu \sim 1 - 10\;\mu$Jy are dominated by actively star-forming galaxies
(e.g., Haarsma et al. 2000).
At $1 \mbox{--} 0.01$~Jy we see an excess of 1.4-GHz sources compared with 
our model calculations.
The excess component mainly consists of radio galaxies that are dominated
by ellipticals and are not directly related to the star formation in galaxies.
But the observed counts and model predictions show a convergence with 
each other toward the fainter flux regime.
This supports the claim of Haarsma et al.\ (2000); thus, we expect that
the millijansky radio sources are really star-forming galaxies.

\begin{figure*}[t]
\centering\includegraphics[width=7cm]{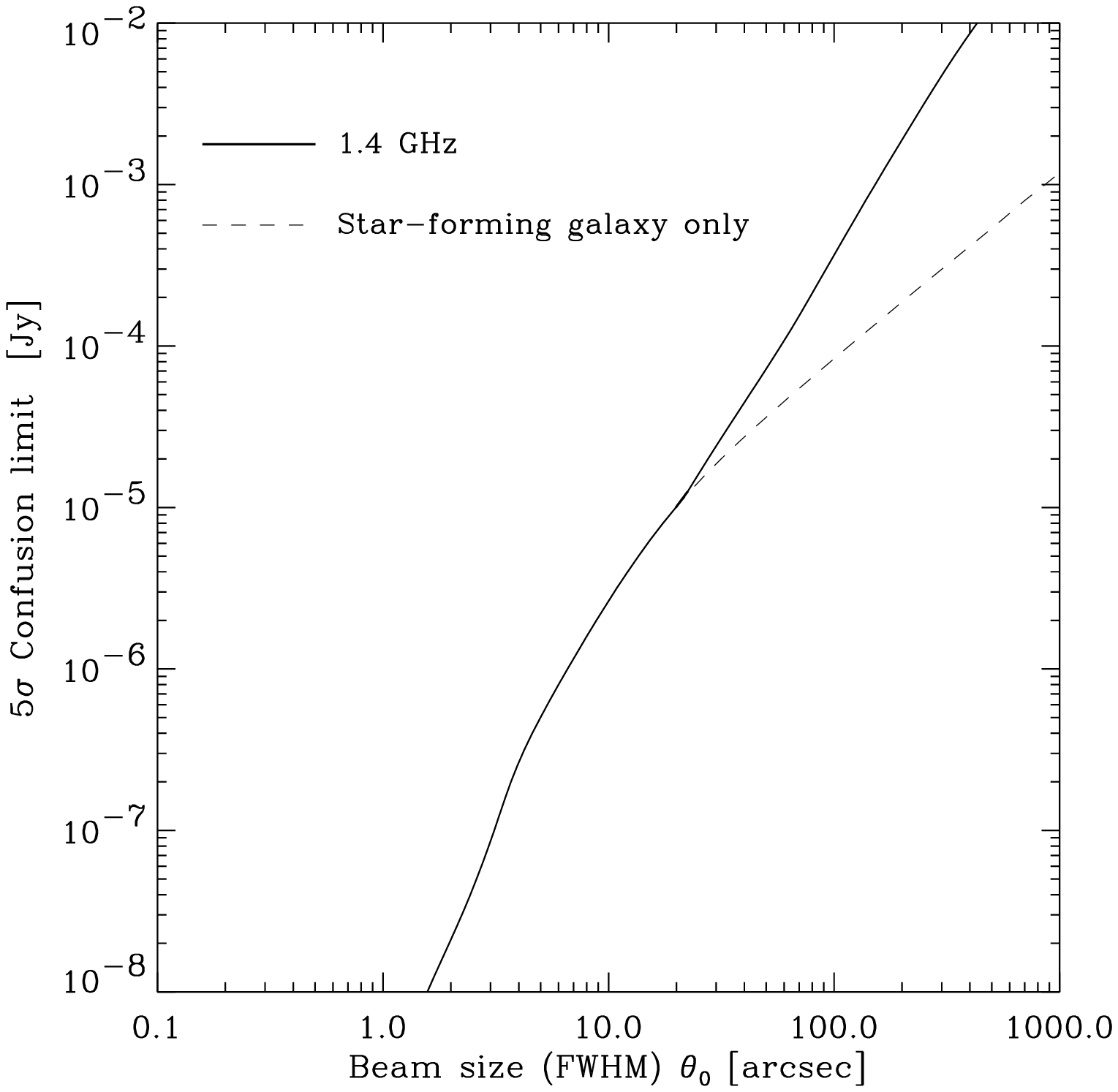}
\figurenum{13}
\figcaption[fig13.ps]{
  The $5\sigma$-source confusion limits as a function of the 
  beam size at 1.4~GHz.
  The same as Figure~\ref{fig:confusion}, we calculated the limit 
  from the Evolution~2 in Figure~\ref{fig:nc}.
  The contribution of nonthermal radio galaxies at the bright flux regime
  is also involved in the calculation.
}\label{fig:confusion_20cm}
\end{figure*}

In such a long wavelength as 1.4~GHz (20 cm), we must be careful of 
the diffraction limit.
We show the $5\sigma$-confusion limit of the 1.4~GHz observations in 
Figure~\ref{fig:confusion_20cm}.
The model basically treats the contribution of the star-forming galaxies,
but as we mentioned above, bright nonthermal sources contribute to the
source confusion at this wavelength.
We take their contribution to the confusion estimation.
The dashed line in Figure~\ref{fig:confusion_20cm} is the confusion
when only the star-forming galaxies are taken into account, and the 
solid thick line represents the confusion limit including strong nonthermal
radio sources.

\begin{figure*}[t]
\centering\includegraphics[width=6cm]{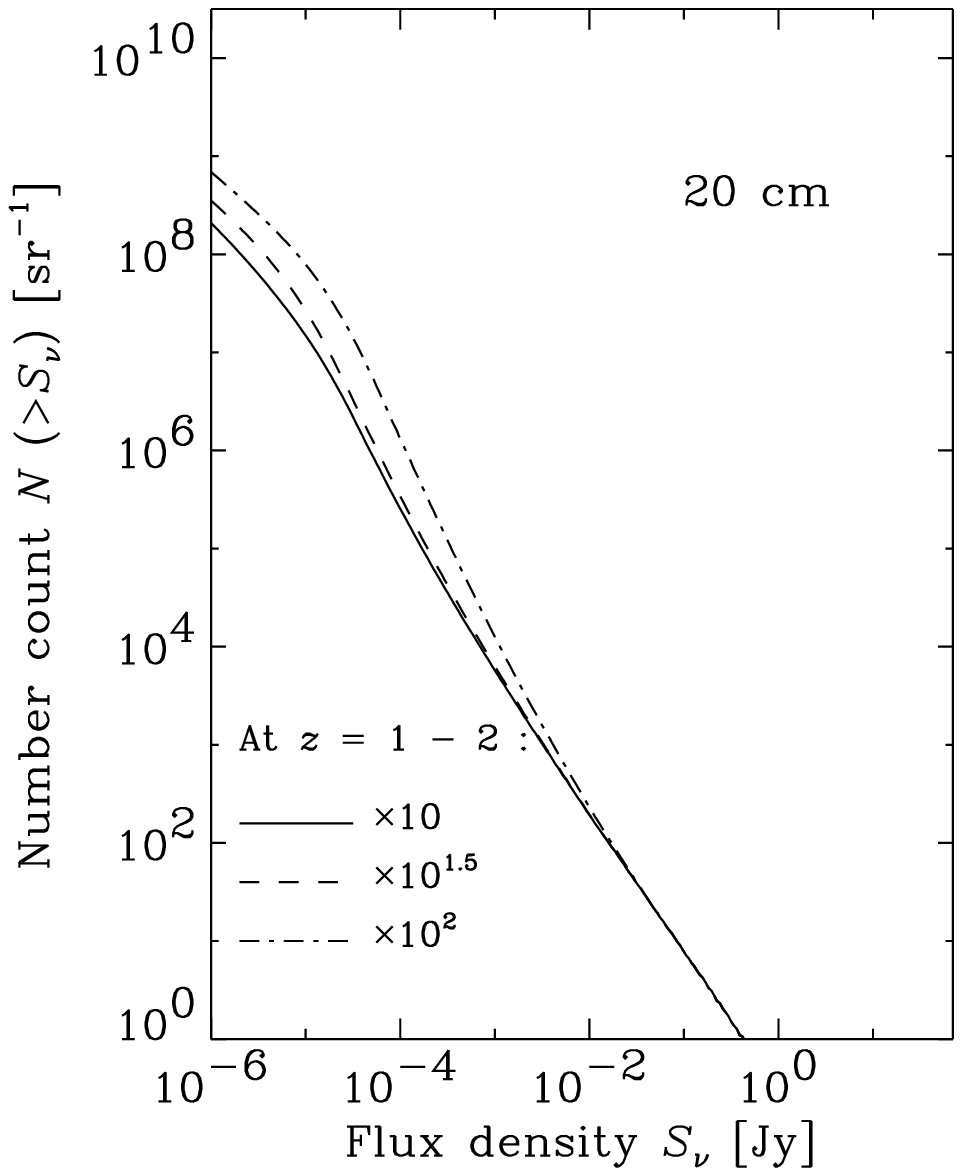}
\figurenum{14}
\figcaption[fig14.ps]{
  The 1.4~GHz source counts based on the evolutionary histories
  shown in the top panel of Figure~\ref{fig:evlp}.
  Same as Figure)\ref{fig:nc_level}, the dotted line shows the no-evolution 
  result, and the solid curve represents the galaxy number counts with 
  Evolution~1.
  The dashed and dot-dashed lines represent the number counts with the
  modified evolutions so that the evolutionary factor at $z = 1 - 2$ 
  is $10^{1.5}$ and $10^{2.0}$, respectively.
  The faint radio counts at 1.4~GHz depend on the evolutionary status of 
  galaxies at $z = 1 \mbox{--} 2$.
}\label{fig:nc_cm_level}
\end{figure*}

\begin{figure*}[t]
\centering\includegraphics[width=10cm]{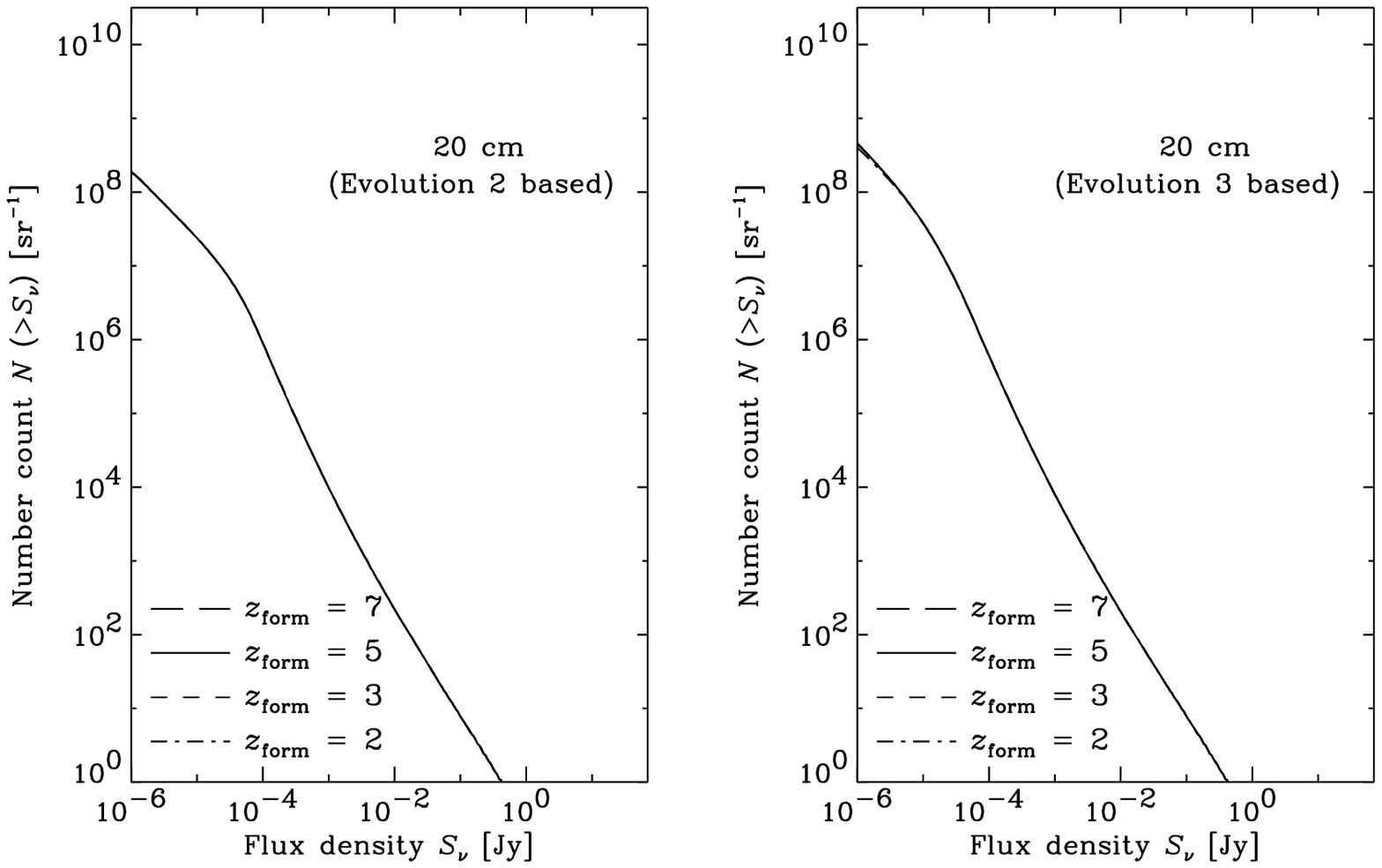}
\figurenum{15}
\figcaption[fig15.ps]{
  The 1.4~GHz source count based with various redshift cutoffs
  $z_{\rm form} = 2, 3, 5$, and 7 based on Evolution~2 and 3 in 
  Figure~\ref{fig:evlp}.
  The dot-dashed, dashed, solid, and long-dashed curves represent the 
  resulting number counts with redshift cutoffs 
  $z_{\rm form} = 2, 3, 5$, and 7, respectively.  
  The number counts with different models are almost completely degenerate, and
  we see that the 1.4-GHz counts are almost insensitive to the evolutionary 
  status of galaxies at $z > 2$.
}\label{fig:nc_cm_zmax}
\end{figure*}

Then we investigate what we know from the faint radio source counts.
We present the radio number counts based on the evolutionary histories
presented in Sections~\ref{subsec:int_z} and \ref{subsec:zmax} in 
Figures~\ref{fig:nc_cm_level} and \ref{fig:nc_cm_zmax}.
Figure~\ref{fig:nc_cm_level} clearly shows that the faint radio counts
at 1.4~GHz depend on the evolutionary status of galaxies at 
$z = 1 \mbox{--} 2$.
On the contrary, Figure~\ref{fig:nc_cm_zmax} demonstrates that 
the 1.4-GHz counts are almost insensitive to the evolutionary status 
of galaxies at $z > 2$.
The redshift information by direct measurement is required for 
very high-$z$ objects to evaluate the star formation activity, 
as studied by Haarsma et al.\ (2000).

This result shows that deep radio surveys are another probe of galaxy 
evolution in the redshift range $z = 1 \mbox{--} 2$.
It should be tested by comparison with future results from submillimeter 
large-area surveys, discussed in Section~\ref{subsec:int_z}.

\section{SUMMARY AND CONCLUSIONS}\label{sec:conclusion}

In this paper we investigated what we can learn about galaxy 
formation and evolution from the data which will be obtained by the forthcoming
new large submillimeter/radio facilities, mainly by the ASTE and the ALMA/LMSA.

We first calculated the source counts from $90\;\mu$m to 3~mm by using
the infrared galaxy number count model of Takeuchi et al.\ (2001).
Based on their number counts, we then derived the source confusion 
noise and estimated the confusion limits at various wavebands 
as a function of the characteristic beam size.

We found that, at the submillimeter wavelengths, source confusion 
for the 10 -- 15-m class facilities becomes severe at the 0.1 to 1~mJy level, 
and astrometry and flux measurement are difficult.
Then we showed that a very large-area survey of the submillimeter sources 
brighter than 10 -- 50~mJy can provide a unique constraint on the infrared 
galaxy evolution at $z = 1 \mbox{--} 2$, and such a survey is suitable 
for a facility such as the ASTE.
A large area is required to suppress the statistical fluctuation caused
by galaxy clustering on the sky.
Such a survey will also enable us to study the clustering properties of the 
bright submillimeter sources, which is still highly unknown.

We also found that the $5\sigma$-confusion limit of LMSA reaches to 
$1 \; \mu$Jy, which enables us to study the contribution of sources 
at extremely large redshifts.
The source counts at such a faint flux level give important information 
on the galaxy formation epoch.
At such faint fluxes the statistical uncertainty stated above 
is small enough that we can safely estimate the galaxy number counts.

We then discussed the possibility of using multiband photometric measurements 
in the infrared (by ASTRO-F in this work) to the millimeter as a rough 
redshift estimator.
We suggested that the source redshift can be obtained from the color and 
flux of the thermal dust emission.
More precise redshift values can be estimated by {\sl `the dust-$z$ method'}, 
which additionally uses the relation between the 
$S_{60\mu{\rm m}}/S_{100\mu{\rm m}}$ color and the luminosity at $60\;\mu$m.
Thus multiband instrument is very welcome for this purpose.
We examined the effectiveness of this method by Monte Carlo simulations and 
found that it successfully works if we have reasonable measurement accuracy.
In addition, the forthcoming IR facilities such as ASTRO-F and {\sl SIRTF} 
will provide very good information of IR SEDs of galaxies, which will 
certainly improve the validity of the method.

We calculated the comparison of the observed 1.4~GHz source counts
and our model counts, to examine the contribution of star forming galaxies
to the faint radio galaxies.
The faint radio counts at 1.4~GHz depend on the evolutionary status of 
galaxies at $z = 1 \mbox{--} 2$ but are insensitive to the evolutionary 
status of galaxies at $z > 2$.
In order to explore the radio properties of such high-$z$ sources, we need 
a direct measurement of their redshifts.

First we thank the anonymous referee for useful suggestions and comments.
We also thank Drs.\ Hiroshi Shibai, T.\ N.\ Rengarajan, Hiroshi Matsuo, 
Hideo Matsuhara, Nobuharu Ukita, and Tomonori Totani for fruitful 
discussions and comments.
We are grateful to Drs.\ Shin Mineshige and Yasushi Suto for 
continuous encouragement.
HH, KY, and KN are grateful to JSPS fellowship.
We made extensive use of the NASA's Astrophysics Data System Abstract 
Service (ADS).

\end{document}